\documentclass[printer]{aa}  
\usepackage{graphicx}
\usepackage{txfonts}
\usepackage{graphicx}
\usepackage{amsmath}
\usepackage{amssymb}
\usepackage{txfonts}
\usepackage{hyperref}
\usepackage{cprotect} % To use verbatim mode in captions
\usepackage{multirow}
\usepackage{textgreek}
\usepackage{subcaption}
\usepackage{booktabs}
\usepackage{caption}
\usepackage{threeparttable}
\usepackage{calrsfs}
\DeclareMathAlphabet{\pazocal}{OMS}{zplm}{m}{n}
\usepackage{adjustbox}

\RequirePackage[normalem]{ulem} %DIF PREAMBLE
\RequirePackage{color}\definecolor{RED}{rgb}{1,0,0}\definecolor{BLUE}{rgb}{0,0,1} %DIF PREAMBLE
\providecommand{\DIFaddbegin}{} %DIF PREAMBLE
\providecommand{\DIFaddend}{} %DIF PREAMBLE
\providecommand{\DIFdelbegin}{} %DIF PREAMBLE
\providecommand{\DIFdelend}{} %DIF PREAMBLE
\providecommand{\DIFaddbeginFL}{} %DIF PREAMBLE
\providecommand{\DIFaddendFL}{} %DIF PREAMBLE
\providecommand{\DIFdelbeginFL}{} %DIF PREAMBLE
\providecommand{\DIFdelendFL}{} %DIF PREAMBLE
\newcommand{\DIFscaledelfig}{0.5}
\RequirePackage{settobox} %DIF PREAMBLE
\RequirePackage{letltxmacro} %DIF PREAMBLE
\newsavebox{\DIFdelgraphicsbox} %DIF PREAMBLE
\newlength{\DIFdelgraphicswidth} %DIF PREAMBLE
\newlength{\DIFdelgraphicsheight} %DIF PREAMBLE
\LetLtxMacro{\DIFOincludegraphics}{\includegraphics} %DIF PREAMBLE
\newcommand{\DIFaddincludegraphics}[2][]{{\color{blue}\fbox{\DIFOincludegraphics[#1]{#2}}}} %DIF PREAMBLE
\newcommand{\DIFdelincludegraphics}[2][]{% %DIF PREAMBLE
\sbox{\DIFdelgraphicsbox}{\DIFOincludegraphics[#1]{#2}}% %DIF PREAMBLE
\settoboxwidth{\DIFdelgraphicswidth}{\DIFdelgraphicsbox} %DIF PREAMBLE
\settoboxtotalheight{\DIFdelgraphicsheight}{\DIFdelgraphicsbox} %DIF PREAMBLE
\scalebox{\DIFscaledelfig}{% %DIF PREAMBLE
\parbox[b]{\DIFdelgraphicswidth}{\usebox{\DIFdelgraphicsbox}\\[-\baselineskip] \rule{\DIFdelgraphicswidth}{0em}}\llap{\resizebox{\DIFdelgraphicswidth}{\DIFdelgraphicsheight}{% %DIF PREAMBLE
\setlength{\unitlength}{\DIFdelgraphicswidth}% %DIF PREAMBLE
\begin{picture}(1,1)% %DIF PREAMBLE
\thicklines\linethickness{2pt} %DIF PREAMBLE
{\color[rgb]{1,0,0}\put(0,0){\framebox(1,1){}}}% %DIF PREAMBLE
{\color[rgb]{1,0,0}\put(0,0){\line( 1,1){1}}}% %DIF PREAMBLE
{\color[rgb]{1,0,0}\put(0,1){\line(1,-1){1}}}% %DIF PREAMBLE
\end{picture}% %DIF PREAMBLE
}\hspace*{3pt}}} %DIF PREAMBLE
} %DIF PREAMBLE
\LetLtxMacro{\DIFOaddbegin}{\DIFaddbegin} %DIF PREAMBLE
\LetLtxMacro{\DIFOaddend}{\DIFaddend} %DIF PREAMBLE
\LetLtxMacro{\DIFOdelbegin}{\DIFdelbegin} %DIF PREAMBLE
\LetLtxMacro{\DIFOdelend}{\DIFdelend} %DIF PREAMBLE
\DeclareRobustCommand{\DIFaddbegin}{\DIFOaddbegin \let\includegraphics\DIFaddincludegraphics} %DIF PREAMBLE
\DeclareRobustCommand{\DIFaddend}{\DIFOaddend \let\includegraphics\DIFOincludegraphics} %DIF PREAMBLE
\DeclareRobustCommand{\DIFdelbegin}{\DIFOdelbegin \let\includegraphics\DIFdelincludegraphics} %DIF PREAMBLE
\DeclareRobustCommand{\DIFdelend}{\DIFOaddend \let\includegraphics\DIFOincludegraphics} %DIF PREAMBLE
\LetLtxMacro{\DIFOaddbeginFL}{\DIFaddbeginFL} %DIF PREAMBLE
\LetLtxMacro{\DIFOaddendFL}{\DIFaddendFL} %DIF PREAMBLE
\LetLtxMacro{\DIFOdelbeginFL}{\DIFdelbeginFL} %DIF PREAMBLE
\LetLtxMacro{\DIFOdelendFL}{\DIFdelendFL} %DIF PREAMBLE
\DeclareRobustCommand{\DIFaddbeginFL}{\DIFOaddbeginFL \let\includegraphics\DIFaddincludegraphics} %DIF PREAMBLE
\DeclareRobustCommand{\DIFaddendFL}{\DIFOaddendFL \let\includegraphics\DIFOincludegraphics} %DIF PREAMBLE
\DeclareRobustCommand{\DIFdelbeginFL}{\DIFOdelbeginFL \let\includegraphics\DIFdelincludegraphics} %DIF PREAMBLE
\DeclareRobustCommand{\DIFdelendFL}{\DIFOaddendFL \let\includegraphics\DIFOincludegraphics} %DIF PREAMBLE
\RequirePackage{listings} %DIF PREAMBLE
\RequirePackage{color} %DIF PREAMBLE
\lstdefinelanguage{DIFcode}{ %DIF PREAMBLE
  moredelim=[il][\color{red}\sout]{\%DIF\ <\ }, %DIF PREAMBLE
  moredelim=[il][\color{blue}\uwave]{\%DIF\ >\ } %DIF PREAMBLE
} %DIF PREAMBLE
\lstdefinestyle{DIFverbatimstyle}{ %DIF PREAMBLE
  language=DIFcode, %DIF PREAMBLE
  basicstyle=\ttfamily, %DIF PREAMBLE
  columns=fullflexible, %DIF PREAMBLE
  keepspaces=true %DIF PREAMBLE
} %DIF PREAMBLE
\lstnewenvironment{DIFverbatim}{\lstset{style=DIFverbatimstyle}}{} %DIF PREAMBLE
\lstnewenvironment{DIFverbatim*}{\lstset{style=DIFverbatimstyle,showspaces=true}}{} %DIF PREAMBLE
\begin{document}

   \title{The MPIfR-MeerKAT Galactic Plane Survey}

   \subtitle{II. The eccentric double neutron star system PSR J1208$-$5936 and a neutron star merger rate update}

   \author{M.~Colom~i~Bernadich\inst{1}
          \and
          V.~Balakrishnan\inst{1}
          \and
          E.~Barr\inst{1}
          \and
          M.~Berezina\inst{1,2}
          \and
          M.~Burgay\inst{3}
          \and
          S.~Buchner\inst{4}
          \and
          D.~J.~Champion\inst{1}
          \and
          W.~Chen\inst{1}
          \and
          G.~Desvignes\inst{1}
          \and
          P.~C.~C.~Freire\inst{1}
          \and
          K.~Grunthal\inst{1}
          \and
          M.~Kramer\inst{1}
          \and
          Y.~Men\inst{1}
          \and
          P.~V.~Padmanabh\inst{5,6,1}
          \and
          A.~Parthasarathy\inst{1}
          \and
          D.~Pillay\inst{1}
          \and
          I.~Rammala\inst{1}
          \and
          S.~Sengupta\inst{1}
          \and
          V.~Venkatraman Krishnan\inst{1}
          }

   \institute{Max-Planck-Institut f\"{u}r Radioastronomie, Auf dem H\"{u}gel 69, D-53121 Bonn, Germany
         \and
             Landessternwarte, Universität Heidelberg, Königstuhl 12, D-69117 Heidelberg, Germany
         \and
             INAF - Osservatorio Astronomico di Cagliari, via della Scienza 5, 09047 Selargius (CA), Italy
         \and
             South African Radio Astronomy Observatory, 2 Fir Street, Black River Park, Observatory 7925, South Africa
         \and
             Max Planck Institute for Gravitational Physics (Albert Einstein Institute), D-30167 Hannover, Germany
         \and
             Leibniz Universit\"{a}t Hannover, D-30167 Hannover, Germany 
             }

   \date{Received September 15, 1996; accepted March 16, 1997}

% \abstract{}{}{}{}{} 
% 5 {} token are mandatory
 
  \abstract
    {The MPIfR-MeerKAT Galactic Plane survey at L-band (MMGPS-L) is the most sensitive pulsar survey in the Southern Hemisphere, providing 78 discoveries in an area of 900 square degrees. Here, we present a follow-up study of one of these new discoveries, PSR J1208$-$5936, a 28.71-ms recycled pulsar in a double neutron star system with an orbital period of $P_\textrm{b}=0.632$ days and an eccentricity of $e=0.348$, merging within the Hubble time. Through timing of almost one year of observations, we detected the relativistic advance of periastron ($\dot\omega=0.918(1)$~deg\,yr$^{-1}$), resulting in a total system mass of $M_\textrm{t}=2.586(5)$~M$_\odot$. We also achieved low-significance constraints on the amplitude of the Einstein delay and Shapiro delay, in turn yielding constraints on the pulsar mass ($M_\textrm{p}=1.26^{+0.13}_{-0.25}$~M$_\odot$), the companion mass ($M_\textrm{c}=1.32^{+0.25}_{-0.13}$~M$_\odot$), and the inclination angle ($i=57\pm12$~degrees). This system is highly eccentric compared to other Galactic field double neutron stars with similar periods, possibly hinting at a larger-than-usual supernova kick during the formation of the second-born neutron star. The binary will merge within $7.2(2)$~Gyr due to the emission of gravitational waves, making it a progenitor of the neutron star merger events seen by ground-based gravitational wave observatories. With the improved sensitivity of the MMGPS-L, we updated the Milky Way neutron star merger rate to be $\pazocal{R}_\textrm{MW}^\textrm{new}=25^{+19}_{-9}$~Myr$^{-1}$ within 90\% credible intervals, which is lower than previous studies based on known Galactic binaries owing to the lack of further detections despite the highly sensitive nature of the survey. This implies a local cosmic neutron star merger rate of $\pazocal{R}_\textrm{local}^\textrm{new}=293^{+222}_{-103}$~Gpc$^{-3}$\,yr$^{-1}$, which is consistent with LIGO and Virgo O3 observations. With this, we also predict the observation of $10^{+8}_{-4}$ neutron star merger events during the LIGO-Virgo-KAGRA O4 run. We predict the uncertainties on the component masses and the inclination angle will be reduced to $5\times10^{-3}$~M$_\odot$ and $0.4$~degrees after two decades of timing, and that in at least a decade from now the detection of $\dot P_\textrm{b}$ and the sky proper motion will serve to make an independent constraint of the distance to the system.}

   \titlerunning{MMGPS II. PSR J1208$-$5936 and a neutron star merger rate update}
   \maketitle
%
%-------------------------------------------------------------------

\section{Introduction}

\begin{table*}
\begin{adjustbox}{max width=\textwidth}
\begin{threeparttable}
\caption[]{\label{DNSsystems} All known pulsars in Galactic DNS systems and candidates with their spin parameters, orbital parameters, and component masses ordered by merger times $\tau_\textrm{m}$, including J1208$-$5936 (highlighted). The first solid line splits systems merging within the Hubble time from those that do not. The last two entries have no total mass measurements, and therefore their DNS nature is uncertain. Nonetheless, they are promising candidates due to their eccentricity.}
\centering
    \begin{tabular}{lccccccccc}
    \hline
    \hline
    PSR & $P_0$ & $\dot{P}$ & $P_\textrm{b}$ & $x$ & $e$ & $M_\textrm{t}$ & $M_\textrm{c}$ & $M_\textrm{p}$ & $\tau_\textrm{m}$ \\ 
     & (ms) & (s\,s$^{-1}$) & (days) & (ls) &  & (M$_\odot$) & (M$_\odot$) & (M$_\odot$) & (Gyr) \\
    \hline
    J1946+2052$^1$ & 16.960 & 9.20$\times10^{-19}$ & 0.078 & 1.154 & 0.064 & 2.50(4) & >1.18 & <1.31 & \textasciitilde0.0455\,$^\textrm{a}$ \\
    J1757$-$1854$^2$ & 21.497 & 2.63$\times10^{-18}$ & 0.184 & 2.238 & 0.606 & 2.732876(8) & 1.3917(4) & 1.3412(4) & 0.0761 \\
    J0737$-$3039A$^3$ & 22.699 & 1.76$\times10^{-18}$ & 0.102 & 1.415 & 0.088 & 2.587052(9) & 1.248868(13) & 1.338185(14) & 0.0860 \\
    J0737$-$3039B$^3$ & 2773.5 & 8.92$\times10^{-16}$ & 0.102 & 1.516 & 0.088 & 2.587052(9) & 1.338185(14) & 1.248868(13) & 0.0860 \\
    B1913+16$^4$ & 59.030 & 8.62$\times10^{-18}$ & 0.323 & 2.342 & 0.617 & 2.828(1) & 1.390(1) & 1.438(1) & 0.301 \\ 
    J1906+0746$^5$ & 144.07 & 2.03$\times10^{-14}$ & 0.166 & 1.42 & 0.085 & 2.6134(3) & 1.322(11) & 1.291(11) & 0.308 \\
    J1913+1102$^6$ & 27.285 & 1.61$\times10^{-19}$ & 0.206 & 1.755 & 0.090 & 2.8887(6) & 1.27(3) & 1.62(3) & 0.470 \\ 
    J0509+3801$^7$ & 76.541 & 7.93$\times10^{-18}$ & 0.38 & 2.051 & 0.586 & 2.805(3) & 1.46(8) & 1.34(8) & 0.576 \\ 
    J1756$-$2251$^8$ & 28.462 & 1.02$\times10^{-18}$ & 0.32 & 2.756 & 0.181 & 2.56999(6) & 1.230(7) & 1.341(7) & 1.66 \\ 
    B1534+12$^9$ & 37.904 & 2.42$\times10^{-18}$ & 0.421 & 3.729 & 0.274 & 2.678463(4) & 1.3455(2) & 1.3330(2) & 2.73 \\ [0.3ex]
    \textbf{J1208}$\mathbf{-}$\textbf{5936}$^{10}$ & \textbf{28.714} & \textbf{<\,4}$\mathbf{\times10^{-20}}$ & \textbf{0.632} & \textbf{4.257} & \textbf{0.348} & \textbf{2.586(6)}~$^\textrm{c}$ & $\mathbf{1.32^{+0.25}_{-0.13}}$~$^\textrm{c}$ & $\mathbf{1.26^{+0.13}_{-0.25}}$~$^\textrm{c}$ & \textbf{7.2(2)} \\

    \hline

    J1829+2456$^{11}$ & 41.010 & 5.25$\times10^{-20}$ & 1.176 & 7.238 & 0.139 & 2.60551(19) & 1.299(4) & 1.306(7) & 55 \\
    J1325$-$6253$^{12}$ & 28.969 & 4.80$\times10^{-20}$ & 1.816 & 7.574 & 0.064 & 2.57(6) & >0.98 & <1.59 & \textasciitilde189\,$^\textrm{a}$ \\
    J1411+2551$^{13}$ & 62.453 & 9.56$\times10^{-20}$ & 2.616 & 9.205 & 0.17 & 2.538(22) & >0.92 & <1.62 & \textasciitilde466\,$^\textrm{a}$ \\ 
    J1759+5036$^{14}$ & 176.02 & 2.43$\times10^{-19}$ & 2.043 & 6.825 & 0.308 & 2.62(3) & >0.7006 & <1.9194 & \textasciitilde177\,$^\textrm{a}$ \\
    J0453+1559$^{15}$ & 45.782 & 1.86$\times10^{-19}$ & 4.072 & 14.467 & 0.113 & 2.734(4) & 1.174(4) & 1.559(5) & 1\,453 \\ 
    J1811$-$1736$^{16}$ & 104.18 & 9.01$\times10^{-19}$ & 18.779 & 34.783 & 0.828 & 2.57(10) & >0.93 & <1.64 & \textasciitilde1\,800\,$^\textrm{a}$ \\ 
    J1518+4904$^{17}$ & 40.935 & 2.72$\times10^{-20}$ & 8.634 & 20.044 & 0.249 & 2.7183(7) & 1.31(8) & 1.41(8) & 8\,844 \\ 
    J1018$-$1523$^{18}$ & 83.152 & 1.09(6)$\times10^{-19}$ & 8.984 & 26.157 & 0.228 & 2.3(3) & >1.16 & <1.1(3) & ~\textasciitilde1.4(3)$\times10^4$\,$^\textrm{(a)}$ \\
    J1930$-$1852$^{19}$ & 185.52 & 1.80$\times10^{-17}$ & 45.06 & 86.89 & 0.399 & 2.59(4) & >1.30 & <1.32 & \textasciitilde5.32$\times10^5$\,$^\textrm{(a)}$ \\ 

    \hline

    J1755$-$2550$^{20}$ & 315.20 & 2.43$\times10^{-15}$ & 9.696 & 12.284 & 0.089 & ... & >0.39 & ... & ...\,$^\textrm{b}$ \\ 
    J1753$-$2240$^{21}$ & 95.138 & 9.70$\times10^{-19}$ & 13.638 & 18.115 & 0.304 & ... & >0.4875 & ... & ...\,$^\textrm{b}$ \\ 

    \hline
    \hline
    \end{tabular}

    \begin{tablenotes}
      \small
      \item $^\textrm{a}$\,Due to uncertain masses, merger times values are only estimates.
      \item $^\textrm{b}$\,Merger time unavailable due to the lack of a total system mass measurement.
      \item $^\textrm{c}$\,Total mass quoted from the direct DDGR fit, component masses quoted from the DDGR $\chi^2$ mapping marginal one-dimensional likelihood distributions, shown in Fig.~\ref{mass_diagram} and explained in Section~\ref{timing}.
      \item References: (1) \cite{stovall2018palfa}, (2) \cite{cameron2023relativistic}, (3) \cite{kramer2021strong}, (4) (\cite{weisberg2016relativistic}), (5) \cite{van2015binary}, (6) \cite{ferdman2020asymmetric}, (7) \cite{lynch2018green}, (8) \cite{ferdman2014psr}, (9) \cite{fonseca2014comprehensive}, (10) this work, (11) \cite{haniewicz2021precise}, (12) \cite{sengar2022high}, (13) \cite{martinez2017pulsar}, (14) \cite{agazie2021green}, (15) \cite{martinez2015pulsar}, (16) \cite{corongiu2007binary}, (17) last published work \citep{janssen2008multi} and section 8.3 of \cite{tauris2017formation}, (18)  \cite{swiggum2023psr}, (19) \cite{swiggum2015psr}, (20) \cite{ng2018psr}, (21) \cite{keith2009psr}
    \end{tablenotes}

\end{threeparttable}
\end{adjustbox}
\end{table*}

Double neutron star (DNS) binaries are the evolutionary endpoint of massive binary systems ($M_\textrm{stars}>8$~M$_\odot$) that survive two supernovae while remaining bound \citep[e.g.][]{tauris2017formation,vigna2018formation,chattopadhyay2020modelling}. In this picture, the primary star undergoes a Type Ib/c core collapse supernova, becoming the first-born neutron star (NS). Subsequently, the system undergoes a common-envelope phase as the secondary evolves out of the main sequence \citep[e.g.][]{van2018high}, after which the outer shells of the evolved secondary are expelled, turning the system into a circular, compact NS - He star binary \citep[e.g.][]{chattopadhyay2020modelling}. As the secondary continues to evolve, a case BB Roche-lobe overflow (RLO) ensues in which the primary NS sustains partial recycling \citep{tauris2015ultra}, burying its magnetic field and spinning up to a rotation period of a few tens of milliseconds \citep{bhattacharya1991formation}. Finally, the system must avoid the disruption during the second, ultra-stripped supernova in which the secondary NS is born with the expulsion of 0.1~to~1~M$_\odot$ from the system, and with a maximum kick velocity of $\approx$$100$~m\,s$^{-1}$ \citep{tauris2015ultra}. The nature of this supernova is still a matter of discussion as it can be triggered by either electron capture in the degenerate ONeMg core or iron core collapse, depending on how much stripping has been suffered by the He star, with more massive remnants undergoing the latter channel and suffering greater mass losses and supernova kicks \citep{tauris2015ultra}. The distinction between these two formation mechanisms is blurry due to the randomness of the resulting orbital parameters arising from the unpredictability of the supernova kick direction and magnitude, but the remaining mass of the second-born NS may be a tell-tale sign, as it has been seen that higher eccentricity systems correlate with more massive second-born NSs and less rapidly spinning recycled pulsars \citep[e.g.][]{sengar2022high,andrews2019double,faulkner2005new,tauris2017formation}. Nonetheless, if all of these steps are completed without disruption, then a new, eccentric DNS system is born.

Observable radio pulsars discovered in DNS systems are amongst the most useful astrophysical tools. With dedicated follow-up and observing campaigns using radio telescopes, their orbital parameters and component masses can be measured to high precision through the technique of pulsar timing \citep[e.g.][]{lorimer2012handbook}, allowing for tests of formation channels and even of fundamental physics. This is demonstrated in the literature on formation mechanisms \citep[e.g.][]{tauris2017formation}, breakthrough tests of gravity \citep[e.g.][]{taylor1979decay,taylor1982new,kramer2021strong}, and the validation or exclusion of dense matter models \citep[e.g.][]{ozel2016masses,hu2020constraining}. Beyond radio observations, the discovery of pulsars in DNS systems has also had an indirect impact on other fields of astrophysics. The observation of the orbital decay of B1913+16 \cite{taylor1979decay} experimentally demonstrated the existence of gravitational waves as predicted by general relativity (GR) for the first time and supported the prediction of its merger in 301~Myr, providing scientific justification for the construction of ground-based gravitational-wave observatories. Indeed, with knowledge of the pulsar DNS population in the Milky Way, their rates of orbital decay, and of the sensitivity of blind surveys on the sky, predictions of the observed cosmic rate of NS mergers can be made \citep[e.g.][]{kim2003probability,kim2010effect,kim2015implications,pol2019future,pol2020updated,grunthal2021revisiting}. The most recent estimate \citep{grunthal2021revisiting} provides an upper limit of  $\pazocal{R}_\textrm{local}\leq597$~yr$^{-1}$, which is consistent with the observed rates at ground-based gravitational wave detectors \citep{abbott2021gwtc}, highlighting the synergy between radio and gravitational wave observations. Finally, the first detection of a gravitational wave signal emitted by the coalescence of two NSs, GW170817 \citep{abbott2017gw170817}, opened a new era for multi-wavelength and multi-messenger astronomy, highlighting DNS systems among other astrophysical systems even further \citep{abbott2017multi}.

However, observable pulsars in DNS systems are are a rarity. At the time of writing only 17 Galactic DNS systems are confirmed, of which only nine are competitive for tests of gravity and NS merger rate predictions (Table~\ref{DNSsystems}). Indeed, simulations predict a Galactic DNS formation rate of just 5--31 Myr$^{-1}$, with only $\sim$0.13\% of massive binaries surviving the DNS formation channels \citep{vigna2018formation}. The number of discoveries is also limited by observational reasons. With the exceptions of J0737$-$3039B, J1906+0746 and perhaps J1755-2550 (Table~\ref{DNSsystems}), typically only the first-born, recycled NS is observable due to their longer-lasting emission and larger beaming fractions, while the second-born NS is set to spin down and cross its pulsar death line a few Myr after birth \citep{lorimer2012handbook}. Limitations are also computational, as the extreme orbital motion of pulsars in compact, relativistic DNS systems hampers the effectiveness of traditional periodicity searches from pulsar surveys due to the Doppler smearing of the signal across an observation \citep{johnston1991detectability}. It is, therefore, necessary to implement computationally expensive and sophisticated pulsar searches to increase the sample of known DNS systems (and hence maximise their science output). These  generally include acceleration searches \citep[e.g.][]{keith2010high}, jerk searches \citep[e.g.][]{andersen2018fourier,eatough2021effelsberg,suresh2022gbt} or even template bank searches \citep[e.g.][]{allen2013einstein,balakrishnan2022coherent}.

The MPIfR-MeerKAT Galactic Plane survey at L-band \citep[MMGPS-L,][]{kramer2016meerkat,stappers2016update,padmanabh2023mmgps-l} has been highly successful in this objective. Designed to discover massive, compact binaries in the southern Galactic plane with a time resampling-based acceleration search, it has yielded the discovery of two previously unknown DNS systems, PSR J1155$-$6529 (Berezina et al. in prep.) and PSR J1208$-$5936 (J1208$-$5936 from now on). Additionally, the MMGPS-L is the most sensitive pulsar survey in the southern sky, making its large coverage area highly relevant to estimate the rate of NS mergers in the Milky Way. In this work, we present the follow-up study of J1208$-$5936, a 28.71-ms recycled pulsar in a highly eccentric orbit with another NS, with whom it is bound to merge in $7.2\pm0.2$~Gyr due to the emission of gravitational waves. In addition, we also use this discovery and the improvement in depth of sky coverage provided by the MMGPS-L to recompute the predicted NS merger rate.

This paper is structured as follows: in Section~\ref{discovery&followup}, we present the discovery and the derivation of an early orbital solution. In Section~\ref{timing}, we show a phase-coherent timing solution and perform preliminary mass measurements with almost one year of timing data. In Section~\ref{properties}, we study its general properties, including its pulse profile, a comparison with other pulsars in DNS systems and a discussion of its formation channel. In Section~\ref{companion_search}, we search for pulsations from its companion. In Section~\ref{merger_implications}, we investigate the implications of the discovery and the performance of the MMGPS-L for estimations of the local NS merger rate, and the detection of these events by ground-based gravitational-wave observatories. And finally, in Section~\ref{biases&prospects} we discuss possible biases and prospects for future improvements.

\section{Discovery and orbital solution}\label{discovery&followup}

Led by the Max Planck Institute for Radioastronomy\footnote{\url{https://www.mpifr-bonn.mpg.de/2169/en}} (MPIfR) in collaboration with the South African Radio Astronomy Observatory\footnote{\url{https://www.sarao.ac.za/}} (SARAO), the MMGPS-L was a fast Fourier transform (FFT)-based pulsar search survey with the MeerKAT radio interferometer\footnote{\url{https://www.sarao.ac.za/science/meerkat/about-meerkat/}} \citep{jonas2016meerkat}. Covering 900 sq. deg. on the southern Galactic plane with the primary aim to discover faint relativistic binaries, the MMGPS-L implemented a time-domain acceleration search \citep{padmanabh2023mmgps-l} with the \texttt{PEASOUP}\footnote{\url{https://github.com/ewanbarr/peasoup}} pipeline \citep{barr2020peasoup}.

J1208$-$5936 was discovered on 30 May 2021 in an observation from May 6th as a topocentric 28.706-ms signal in the FFT with $S/N=15.1$, a line-of-sight acceleration of $-8.86$~m\,s$^{-2}$, $\textrm{DM}=344.2$~pc\,cm$^{-3}$ and a \texttt{PulsarX}\footnote{\url{https://github.com/ypmen/PulsarX}} fold of $S/N\approx19$. Upon discovery, we performed a first 20 minute-long follow-up observation at the same position on 4 June 2021. This yielded a re-detection with $S/N\approx15$, and a \texttt{PRESTO/prepfold}\footnote{\url{https://github.com/scottransom/presto}} fold of $S/N\approx12$ with a changed period of $P=28.697$~ms, thus confirming both its existence and its binary nature.

The interferometric nature of MeerKAT, and therefore the multibeam pattern of the follow-up observation allowed for precise localisation of J1208$-$5936 with the \texttt{SeeKAT}\footnote{\url{https://github.com/BezuidenhoutMC/SeeKAT}} software (Bezuidenhout et al. in prep.), which compares the $S/N$ of detections in different beams and does a maximum-likelihood estimate of the best position of the source, taking into account beam positions and their point-spread functions as derived by the \texttt{Mosaic}\footnote{\url{https://github.com/wchenastro/Mosaic}} software \citep{chen2021wide}. This improved the sky position to an uncertainty of just $\sim$5 arcsec, much better than the $\sim$20 arcsec precision provided by the beam size alone, and which was only 3.3 arcsec away from the timing-derived position (Table~\ref{measurements}).

\begin{figure}
\centering
 \includegraphics[width=\columnwidth]{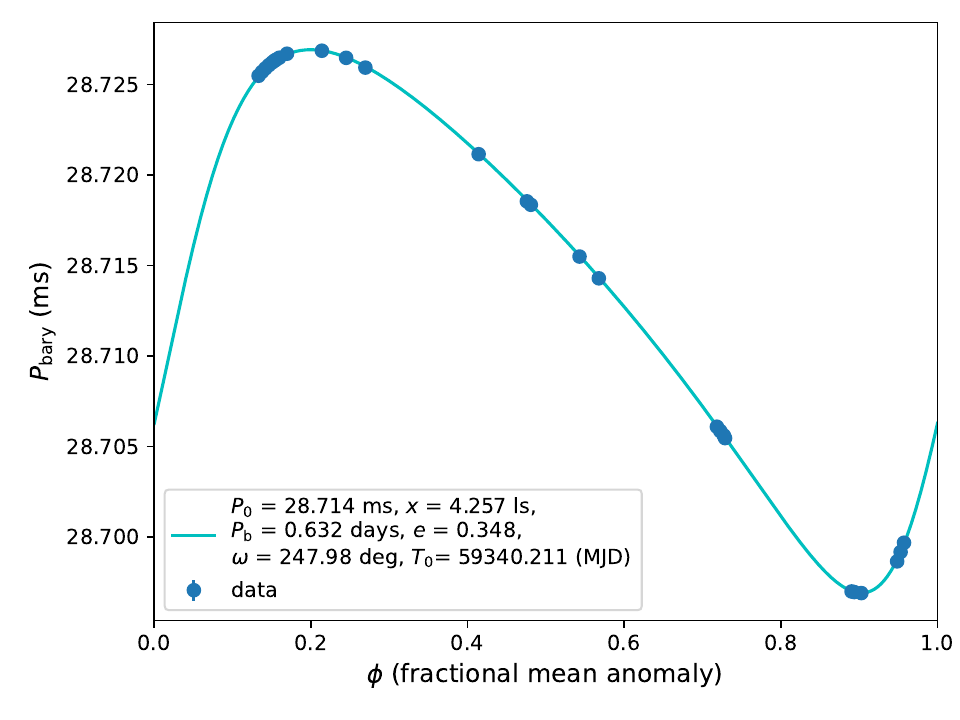}
 \caption{Orbital solution of J1208$-$5936 based on the $P_\textrm{bary}(t_\textrm{obs})$ time series between 5 May 2021 and 31 August 2021. Observations with precise acceleration and/or jerk measurements were used to extract several data points (5 May: $-10.5$~m\,s$^{-2}$; 11 June: $+22.4$~m\,s$^{-2}$, and 16 August: $+7.2$~m\,s$^{-2}$, $-3.6\times10^{-3}$~m\,s$^{-3}$; at $\phi\approx0.72$, $0.95$ and $0.15$). Assuming a canonical pulsar mass of $M_\textrm{p}=1.35$~M$_\odot$, the minimum companion mass computed from the mass function is $M_\textrm{c}\gtrsim1.1$~M$_\odot$ (see Table~\ref{measurements}).}
 \label{orbital_solution}
\end{figure}

From 4 June 2021 to 31 August 2021, we typically scheduled dedicated sessions twice a week, consisting of two 20-minute observations stored as filterbank data in the accelerated pulsar search user supplied equipment cluster in South Africa \citep[APSUSE-mode,][]{padmanabh2023mmgps-l}. Each session had the two observations conducted just a few hours apart. We folded each observation into archives with the \texttt{dspsr}\footnote{\url{https://dspsr.sourceforge.net/}}, where cycles of the pulsar are stacked every 8 or 10 seconds. We then cleaned these archives with the \texttt{clfd}\footnote{\url{https://github.com/v-morello/clfd}} radio-frequency interference (RFI) excision software \citep{morello2019high}. These resulting data allow for accurate tracking of the evolution or drift of the pulse over time, leading to measurements of the barycentric period ($P_\textrm{bary}$) at each observing epoch ($t_\textrm{obs}$) with the \texttt{PSRCHIVE/pdmp}\footnote{\url{http://psrchive.sourceforge.net/}} software \citep{hotan2004psrchive}, but also with the timing software \texttt{TEMPO2}\footnote{\url{https://bitbucket.org/psrsoft/tempo2}} \citep{hobbs2006tempo2,edwards2006tempo2} for observations with large line-of-sight acceleration or jerk.

We implemented a modified version of the roughness estimate algorithm (REA) from \cite{bhattacharyya2008determination} to solve for the orbital period, available in \texttt{estimateOrbit.py}\footnote{\url{https://github.com/mcbernadich/CandyCracker/blob/main/estimateOrbit.py}}. Given adequate orbital coverage, the REA is more sensitive to orbital periods in systems with non-zero eccentricity than the Lomb-Scargle periodogram. It performs a search in the orbital period $P_\textrm{b}$ space by folding the $P_\textrm{bary}(t_\textrm{obs})$ series and evaluating the smoothness of the resulting curve with
\begin{equation}R=\sum^{n}_{i=1}\left[\frac{P_\textrm{bary}(t_{i+1})-P_\textrm{bary}(t_{i})}{\phi(t_{i+1})-\phi(t_{i})}\right]^2\,\textrm{,}\end{equation}
where $t_i=t_{\textrm{obs,}i}/P_\textrm{b,trial}-\textrm{mod}(t_{\textrm{obs,}i},P_\textrm{b,trial})$ are the folded observation epochs and $\phi$ corresponds to the orbital phase (mean anomaly). We then select $P_\textrm{b,trial}$ values that minimise $R$, as they produce a smooth fold and are likely to correspond to the true $P_\textrm{b}$ value. The normalisation by the difference in orbital phase $\Delta\phi$ is a modification of the original REA presented in \cite{bhattacharyya2008determination} that gives higher significance to data points that are closer in orbital phase, dealing with possible gaps in the orbital coverage.

The REA found a significant signal at $P_\textrm{b,trial}=0.632$~days. We then used \texttt{pyfitorbit}\footnote{\url{https://github.com/gdesvignes/pyfitorbit}} software to fit for the remaining parameters using the best REA $P_\textrm{b,trial}$ as a first guess, which resulted in a confirmation of the orbital period and a fit for the remaining Keplerian parameters (Fig.~\ref{orbital_solution}), with a eccentricity $e=0.34$ and a mass function $f_M=0.208$~M$_\odot$.

\section{Timing and mass measurements}\label{timing}

\begin{figure*}
\centering
 \includegraphics[width=2\columnwidth]{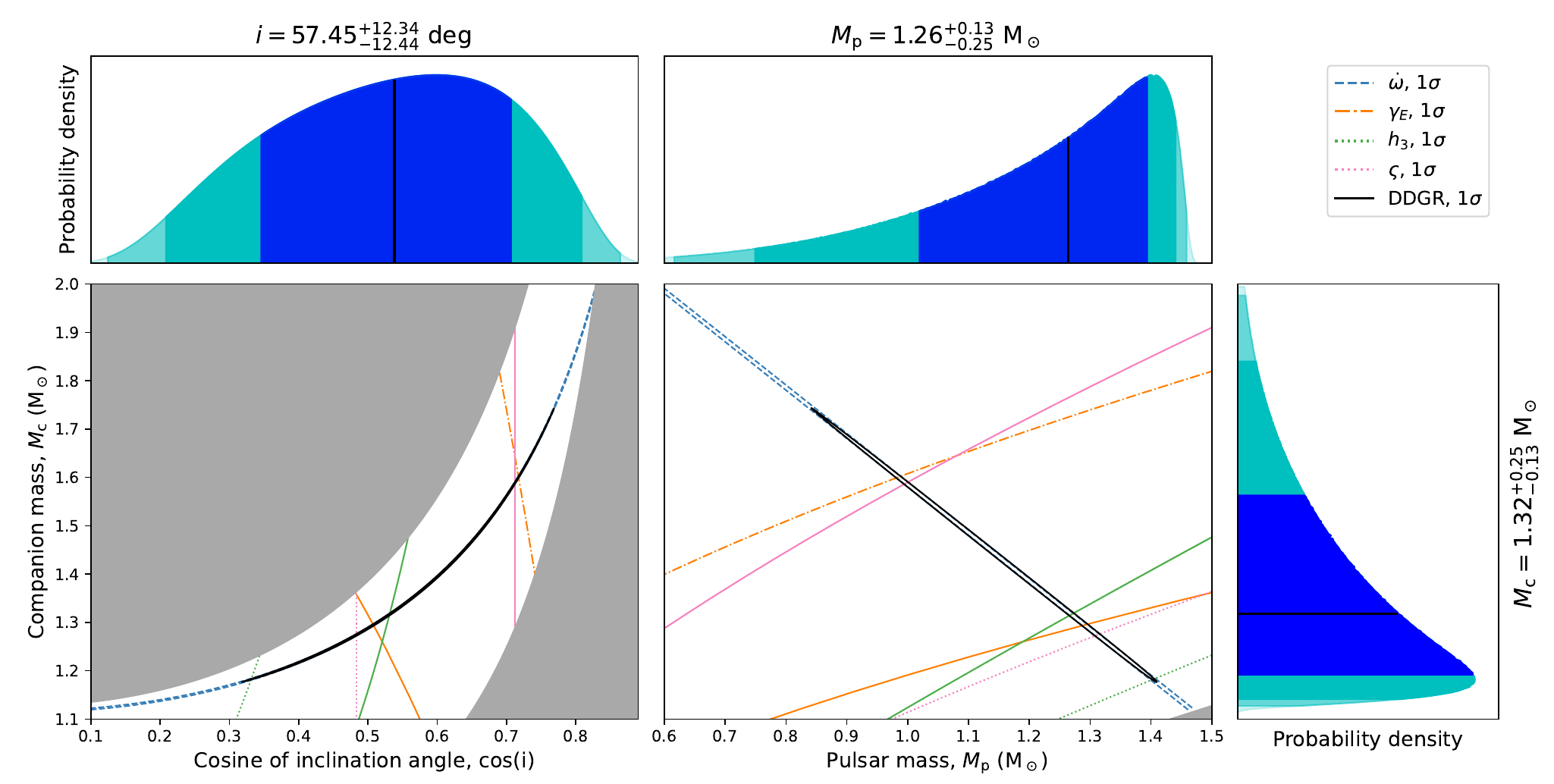}
 \caption{Corner plot showing the constraints on the inclination angle and component masses from PK parameters and the $\chi^2$ mapping of DDGR solutions. \textbf{Central plots}: Mass-mass diagrams portraying the DDH PK parameters constraints, with each color corresponding to a different parameter as indicated in the legend (solid lines: nominal value, dashed lines: 1\textsigma~limits), and the 1\textsigma~limits 
 from DDGR $\chi^2$ mapping in black. The shaded grey area on the right plot represents the region excluded by the mass function ($i>90$), while the shaded areas on the left plot represent the areas outside the prior for $M_\textrm{p}$ (outside of $0.6<M_\textrm{p}<1.5$~M$_\odot$). The explored regions of the $M_\textrm{p}$ and $M_\textrm{c}$ were decided based on the limits given by the mass function and $M_\textrm{t}$ (from $\dot\omega$). \textbf{Outer plots}: marginalised one-dimensional probability densities for $M_\textrm{p}$, $M_\textrm{c}$ and $\cos{i}$ from DDGR $\chi^2$ mapping, showcasing the median value (black solid line) and the 31.4\%, 47.4\% and 49.9\% percentiles on both sides (shaded areas). The resulting mass constraints are consistent with a pair of NSs.}
 \label{mass_diagram}
\end{figure*}

After August 2021, follow-up observations were scheduled to focus on the timing analysis. The observing time was dropped to monthly 5-min-long observations, but in exchange for the reduced observational time, observations were moved to being recorded with the dedicated MeerKAT pulsar timing user supplied equipment \citep[PTUSE,][]{bailes2020meerkat} and APSUSE simultaneously. The advantages of PTUSE over APSUSE are GPU-based coherent de-dispersion, recording of full-Stokes information, real-time data folding and a finer sampling resolution of $9$~\textmu s at the same radio band, thus enabling high precision pulsar timing.

To achieve a phase-coherent timing solution, we used \texttt{dracula2.py}\footnote{\url{https://github.com/mcbernadich/CandyCracker/blob/main/dracula2.py}}, an implementation of the \textit{dracula} algorithm \citep{freire2018algorithm} with \texttt{python} and \texttt{TEMPO2}. \textit{dracula} searches for timing solutions assuming different combinations of phase wraps in between the observations, finding the unique solution that accounts for all the rotations of the pulsar from start to end.
For this, we used PTUSE pulsar archives from the newer observations and APSUSE archives produced from older search data with \texttt{dspsr}, both of them with 128 phase bins across the profile. Those archives were de-dispersed and scrunched in frequency with the \texttt{PSRCHIVE/pam} command, and three Times of Arrival (ToAs) were produced per observation with \texttt{PSRCHIVE/paas}, using a single, narrow von Mises function fitted with \texttt{PSRCHIVE/pat} as an initial timing template. \texttt{dracula.py} quickly converged into a single solution without the need of fitting for the sky position in the process, indicative of the quality of the \texttt{SeeKAT} multibeam localisation.

Given the massive, eccentric and compact nature of the system, we performed a semi-coherent orbital campaign with 14 observations at selected orbital phases from 2 March 2022 to 7 March 2022, accumulating a total of 11 hours.
Using our phase-connected timing model as a predictor, the orbital phases of the observations were chosen to cover features of the unabsorbed signal of a potential Shapiro delay signal \citep{freire2010orthometric} and the passage of periastron.

Together with an additional observation from 8 April 2022, the added PTUSE folded pulse profile with 1024 bins presents $S/N\approx200$, while the added 512-bin APSUSE profile presents $S/N\approx170$, the main difference in $S/N$ coming from the improvement provided by coherent de-dispersion of PTUSE data. These two high $S/N$ profiles were scrunched into four frequency sub-bands for frequency-resolved timing of the entire data set in order to account for DM evolution. Frequency-resolved analytical timing templates were produced with \texttt{paas} both for the PTUSE and APSUSE data sets. Subsequently and with \texttt{dspsr}, all PTUSE data were folded into 1024-bin, four-channel archives, and APSUSE data (when PTUSE data were not available) were folded into 512-bin, four-channel archives, with sub-integration times of at most 10 minutes in length in both cases. These files had been excised of RFI a priori at full time and frequency resolution using first the \texttt{clfd}\footnote{\url{https://github.com/v-morello/clfd}} software \citep{morello2019high} and then manually with \texttt{PSRCHIVE/paz} upon inspection of the remaining RFI. Finally, \texttt{pat} was used with the Fourier Phase Gradient algorithm \citep{taylor1992pulsar} to generate 384 frequency-resolved ToAs.

Assuming GR to be the correct theory of gravity, we analysed the produced ToAs to constrain the pulsar mass ($M_\textrm{p}$), the companion mass ($M_\textrm{c}$) and the inclination angle of the orbit ($i$) through two different methods. The first one is by measuring the theory-independent post-Keplerian (PK) parameters that arise from relativistic corrections of the orbital motion, and then deriving probability densities for $M_\textrm{p}$, $M_\textrm{c}$ and $i$ on which we quote median values and 1\textsigma~uncertainties from the two adjacent 34.1\% percentiles (68.2\% credible intervals). In this work, we measure the advance of periastron passage ($\dot\omega$), the amplitude of the Einstein delay ($\gamma_\textrm{E}$), and the orthometric parameters of Shapiro delay: the third-order orthometric amplitude ($h_3$) and the orthometric ratio ($\varsigma$) as defined in \cite{freire2010orthometric}. In the first post-Newtonian approximation of GR, $\dot\omega$ arises from the rotation of the Keplerian ellipse of the orbit in the direction of the orbital motion, depending on the total mass $M_\textrm{t}=M_\textrm{p}+M_\textrm{c}$ as
\begin{equation}\label{omdot}\dot\omega=3\left(\frac{G}{c^3}\right)^{2/3}\left(\frac{P_\textrm{b}}{2\pi}\right)^{-5/3}\frac{M_\textrm{t}^{2/3}}{1-e^2}\textrm{,}\end{equation}
while $h_3$ and $\varsigma$ parameterise the unabsorbed part of the Shapiro delay \citep{freire2010orthometric}. These describe the delay of the pulses propagating through the companion's gravitational field, depending on $M_\textrm{c}$ and $i$ as
\begin{equation}\label{h3}h_3=\frac{GM_\textrm{c}}{c^3}\left(\frac{1-\cos{i}}{1+\cos{i}}\right)^{3/2}\end{equation}
and
\begin{equation}\label{stig}\varsigma=\left(\frac{1-\cos{i}}{1+\cos{i}}\right)^{1/2}\textrm{,}\end{equation}
for $\cos{i}>0$, which is applicable to our study as we do not have any constrain on the angle of the ascending node ($\Omega_A$, see Section~\ref{prospects_estimation} for a discussion on this). It is worth noting that we use the $h_3$ and $\varsigma$ parametrisation introduced in \cite{freire2010orthometric} instead of the classic range ($r$) and shape ($s$) parameters used in other works \citep[e.g.][]{kramer2021strong} because they better describe systems with low $i$ and are less correlated with each other. And finally, the Einstein delay is caused by the periodic modulation of the relativistic time dilation due to the pulsar's changing orbital velocity and its motion  across the companion's gravitational field, its amplitude $\gamma_\textrm{E}$ being modelled by
\begin{equation}\label{gamma}\gamma_\textrm{E}=\left(\frac{G}{c^3}\right)^{2/3}\left(\frac{P_\textrm{b}}{2\pi}\right)^{1/3}\frac{M_\textrm{c}\left(M_\textrm{c}+M_\textrm{t}\right)}{M_\textrm{t}^{4/3}}e\textrm{.}\end{equation}
These parameters are implemented in DDH, which is an extended version of the Damour-Deruelle pulsar timing model \citep[DD][]{damour1986general} that implements the orthometric parameters $h_3$ and $\varsigma$ instead of $r$ and $s$. The fit was done with \texttt{TEMPO2/TempoNEST}\footnote{\url{https://github.com/LindleyLentati/TempoNest}}, a multi-nested Bayesian sampling plug-in to \texttt{TEMPO2} \citep{lentati2014temponest}, in order to find stable values for both $h_3$ and $\varsigma$ with realistic uncertainties.

The other method is to assume GR from the start with the DDGR model \citep{taylor1989further}, which measures $M_\textrm{t}$ and $M_\textrm{c}$ directly to model the PK effects. For this measurement we implemented a common likelihood approach \citep[first introduced by][see therein for more details]{splaver2002masses} with \texttt{chi2Map.py}\footnote{\url{https://github.com/mcbernadich/mass-diagrams}}, computing the likely constraints from the quality of the \texttt{TEMPO2} fits of the DDGR model in a uniformly spaced grid on the $M_\textrm{t} - \cos{i}$ plane, which produces agnostic prior that follows the random distribution of $i$ values of binary systems in the sky.

The resulting probability distributions, derived from multiplying the measured PK parameters from DDH and the $\chi^2$ values from DDGR, were then marginalised into one-dimensional probability densities for $M_\textrm{p}$, $M_\textrm{c}$ and $\cos{i}$, on which we quote median values and 1\textsigma~uncertainties from the two adjacent 34.1\% percentiles (68.2\% credible intervals). The resulting \texttt{TempoNEST} fit of the DDH model and the \texttt{chi2Map.py} exploration of the DDGR $\chi^2$ space are consistent with each other. The most significant PK parameter is $\dot\omega$, detected in DDH with $\approx$900\textsigma~significance, yielding a highly precise measurement of the total system mass at 
\begin{equation}M^\textrm{DDH}_\textrm{t}=2.586\pm{0.005}~\textrm{M}_\odot\textrm{.}\end{equation}
This is in excellent agreement with the
\begin{equation}M^\textrm{DDGR}_\textrm{t}=2.586\pm{0.006}~\textrm{M}_\odot\end{equation}
given by the direct DDGR \texttt{TEMPO2} fit. We note that $\varsigma$ and $\gamma_\textrm{E}$ are detected with low significance, and there is an important upper limit of Shapiro delay amplitude $h_3$. The derived 1\textsigma~constraints from each DDH PK parameter and the $\chi^2$ DDGR fits are overlayed in Fig.~\ref{mass_diagram}, with the DDGR contour tracing very strictly the limits imposed by $\dot\omega$, and being consistent with the loose limits imposed by $\gamma_\textrm{E}$, $h_3$ and $\varsigma$. The most likely $M_\textrm{p}$, $M_\textrm{c}$ and $i$ values derived from the DDH and DDGR models are in very good consistency with each other, with 
\begin{equation}
\begin{aligned}
& M^\textrm{DDH}_\textrm{p}=1.24^{+0.12}_{-0.18}~\textrm{M}_\odot\textrm{,}\\
& M^\textrm{DDH}_\textrm{c}=1.36^{+0.18}_{-0.12}~\textrm{M}_\odot\textrm{, and}\\
& i^\textrm{DDH}=59\pm9~\textrm{deg}
\end{aligned}
\end{equation}
from multiplying probability distributions given by the PK parameter limits, and
\begin{equation}
\begin{aligned}
& M^\textrm{DDGR}_\textrm{p}=1.26^{+0.13}_{-0.25}~\textrm{M}_\odot\textrm{,}\\
& M^\textrm{DDGR}_\textrm{c}=1.32^{+0.25}_{-0.13}~\textrm{M}_\odot\textrm{, and}\\
& i^\textrm{DDGR}=57\pm12~\textrm{deg}
\end{aligned}
\end{equation}
from the $\chi^2$ mapping of DDGR. Additionally, as a final consistency check, we also re-derived the likely values of $\gamma_\textrm{E}$, $h_3$ and $\varsigma$ by calculating their marginal one-dimensional probability densities from the DDGR $\chi^2$ mapping experiment, on which we quote median values and 1\textsigma~uncertainties from the two adjacent 34.1\% percentiles, and attest that they are in good consistency with their DDH limits (Table~\ref{measurements}).

\begin{figure*}
\centering
 \includegraphics[width=1.75\columnwidth]{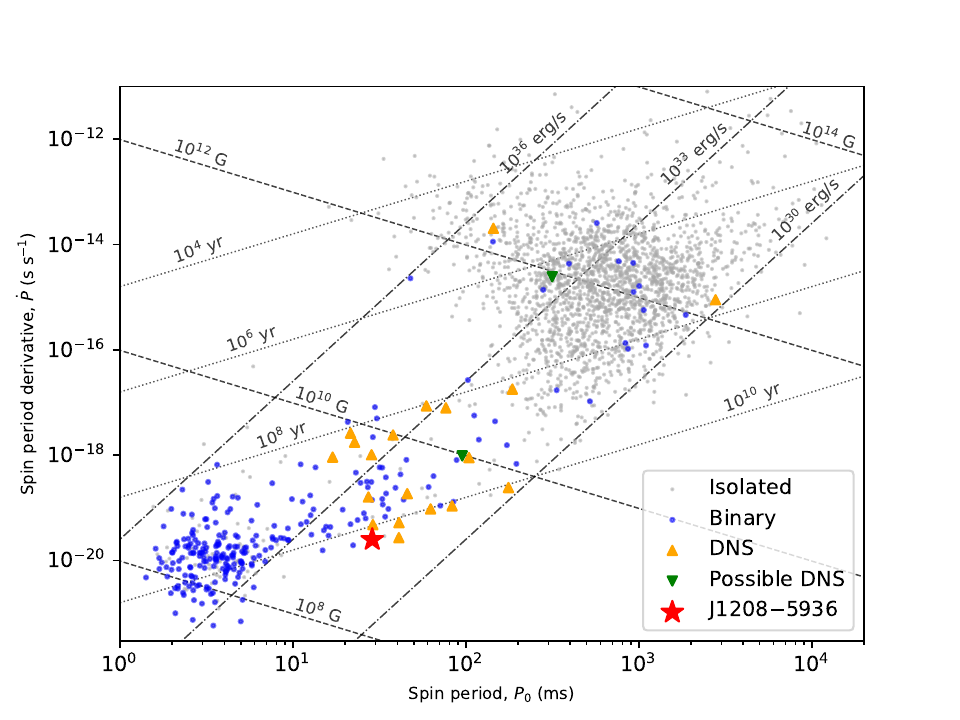}
 \caption{$P_0 - \dot P$ diagram showcasing all isolated and binary pulsars, and highlighting all known DNS systems and candidates (Table~\ref{DNSsystems}). The dashed lines indicate the characteristic age $\tau_\textrm{c}$, the surface magnetic field $B_\textrm{surf}$ and the spin-down luminosity. J1208$-$5936 falls at the bottom of the mildly recycled population. All data points except J1208$-$5936 have been retrieved from the Australia Telescope National Facility (ATNF) website.}
 \label{p-pdot}
\end{figure*}

We also attempted to fit any orbital period derivative $\dot P_\textrm{b}$ or apparent change of the projected semi-major axis $\dot x$, but none of them yield significant limits with DDH. From the DDGR $\chi^2$ mapping limits on $M_\textrm{p}$, $M_\textrm{c}$ and $i$ and assuming GR, we predict them to have values of $\dot P_\textrm{b}^\textrm{GR}=-1.225^{+0.026}_{-0.009}\times10^{-13}$~s\,s$^{-1}$ and $\dot x^\textrm{GR}=-6.37^{+0.14}_{-0.05}\times10^{-18}$~ls\,s$^{-1}$, but the current timing sensitivity is not high enough to yield a proper measurement. Nonetheless, $\dot P_\textrm{b}$ is not likely to be detected without the addition of another decade of timing and it may be contaminated by effects introduced by the galactic acceleration field and proper motion in the sky, while $\dot x$ is likely to be dominated by proper motion effects (Section~\ref{biases_section}, Table~\ref{biases}).

\section{Properties of J1208-5936}\label{properties}

\subsection{General characteristics}

\begin{table*}
\begin{adjustbox}{max width=\textwidth}
\begin{threeparttable}
\caption[]{\label{measurements} Timing parameters resulting from  the DDH and DDGR fits. Values without brackets come from the \texttt{TempoNEST} fit of DDH or the direct \texttt{TEMPO2} fit of DDGR. Values with square brackets [...] are derived from the \texttt{TempoNEST} DDH parameters or the direct \texttt{TEMPO2} DDGR parameters ($M_\textrm{t}$ for DDH and $\dot\omega$ for DDGR). Values with curly brackets \{...\} are derived from the multiplication of PK density probabilities in DDH or the $\chi^2$ mapping probability in DDGR. All Keplerian, spin, position and DM parameters presented for DDGR are taken from the direct \texttt{TEMPO2} fit. These fits also include a time jump between the APSUSE and PTUSE data-sets, as well as a DM jump (see Appendix~\ref{DMjump} for details).}
\centering
\begin{tabular}{lcc}
\hline
\hline
Data reduction parameters & & \\
\hline
Binary model & DDH & DDGR \\
\texttt{TEMPO2} wrapper & \texttt{TempoNEST} & \texttt{chi2Map.py} \\
Solar System ephemeris & DE430 & DE430 \\
Timescale & TCB & TCB \\
Reference epoch for period and DM & 59390 & 59390 \\
Number of ToAs & 384 & 384 \\
Root mean squared of ToA residuals (\textmu s) & 36.99 & 36.93\,$^\textrm{a}$ \\
Reduced $\chi^2$ & 0.949 & 0.944\,$^\textrm{a}$ \\
\hline
Spin and astrometric parameters &  &  \\
\hline
Right ascension, $\alpha$ (J2000, hh:mm:ss) & 12:08:27.024(1) & 12:08:27.023(1) \\
Declination, $\delta$ (J2000, dd:mm:ss) & -59:36:20.485(5) & -59:36:20.486(6) \\
Spin frequency, $\nu$ (Hz) & 34.8263871091(1) & 34.8263871091(2) \\
Spin-down rate, $\dot\nu$ ($10^{-17}$~Hz\,s$^{-1}$) & -3.6(1.1) & -3.1(1.3) \\
Dispersion measure, DM (pc~cm$^{-3}$) & 344.427(6) & 344.428(6) \\
First derivative of DM, DM1 ($10^{-2}$~pc~cm$^{-3}$~yr$^{-1}$) & 2.36(86) & 2.40(98) \\
\hline
Keplerian orbital parameters &  &  \\
\hline
Orbital period, $P_B$ (days) & 0.631566177(3) & 0.631566176(4) \\
Orbital eccentricity, $e$ & 0.347988(1) & 0.347988(2) \\
Longitude of periastron, $\omega$ (deg) & 247.98(1) & 247.98(2) \\
Projected semi-major axis of the pulsar orbit, $x$ (ls) & 4.2570(4) & 4.2571(5) \\
Epoch of periastron, $T_0$ (MJD) & 59340.210571(2) & 59340.210572(2) \\
\hline
Post-Keplerian (PK) orbital parameters &  &  \\
\hline
Rate of advance of periastron $\dot\omega$ (deg\,yr$^{-1}$) & 0.918(1) & [0.918(1)]\,$^\textrm{b}$ \\
Amplitude of Einstein delay, $\gamma_\textrm{E}$ (ms) & 2.93(98) & \{$3.01^{+0.80}_{-0.40}$\} \\[.3ex]
Orthometric amplitude of Shapiro delay, $h_3$ (\textmu s) & 1.10(97) & \{$1.06^{+0.92}_{-0.52}$\} \\[.3ex]
Orthometric ratio of Shapiro delay, $\varsigma$ & 0.41(18) & \{$0.55^{+0.15}_{-0.13}$\} \\[.3ex]
Derivative of orbital period, $\dot P_\textrm{b}^\textrm{GR}$ ($10^{-13}$~s\,s$^{-1}$) & ... & \{$-1.225^{+0.026}_{-0.009}$\} \\[.3ex]
Derivative of projected semi-major axis, $\dot x^\textrm{GR}$ ($10^{-18}$~ls\,s$^{-1}$) & ... & \{$-6.37^{+0.14}_{-0.05}$\} \\
\hline
Mass measurements and derived orbital parameters &  &  \\
\hline
Mass function, $f_M$ (M$_\odot$) & 0.20758(6) & 0.20760(7) \\
Total mass, $M_\textrm{t}$ (M$_\odot$) & [2.586(5)]\,$^\textrm{b}$ & 2.586(6) \\
Companion mass, $M_\textrm{c}$ (M$_\odot$) & \{$1.36^{+0.18}_{-0.12}$\} & 1.39(37)~/~\{$1.32^{+0.25}_{-0.13}$\}\,$^\textrm{c}$ \\[0.3ex]
Pulsar mass, $M_\textrm{p}$ (M$_\odot$) & \{$1.24^{+0.12}_{-0.18}$\} & \{$1.26^{+0.13}_{-0.25}$\} \\[0.3ex]
Inclination angle, $i$ (deg) & \{$55(9)$\} & \{$57(12)$\} \\
Merger time, $\tau_\textrm{m}$ (Gyr) & \multicolumn{2}{c}{7.2(2)} \\
\hline
Derived spin and astrometric parameters & & \\
\hline
Galactic longitude, $l$ (deg) & \multicolumn{2}{c}{297.512} \\
Galactic latitude, $b$ (deg) & \multicolumn{2}{c}{2.813} \\
Spin period, $P_0$ (ms) & 28.7138598921(1) & 28.7138598921(2) \\
Spin period derivative, $\dot{P}$ ($10^{-20}$~s\,s$^{-1}$) & 2.98(0.90) & 2.57(1.04) \\
Characteristic age, $\tau_\textrm{c}$ (Gyr) & \multicolumn{2}{c}{>10} \\
Surface magnetic field strength, $B_\textrm{surf}$ ($10^9$~G) & \multicolumn{2}{c}{<30}\\
NE2001 DM-derived distance, $d$ (kpc) & \multicolumn{2}{c}{8.2(1.6)} \\
YMW16 DM-derived distance, $d$ (kpc) &  \multicolumn{2}{c}{8.5(1.7)} \\
\hline
\hline
\end{tabular}

\begin{tablenotes}
      \small
      \item $^\textrm{a}$ Taken from the direct \texttt{TEMPO2} DDGR fit. Variations from the $\chi^2$ mapping (Section~\ref{timing}) are only of the order of $1/369\approx0.003$ (Reduced $\chi^2=\chi^2/$ degrees of freedom).
      \item $^\textrm{b}$ Derived through equation~\ref{omdot}. In the DDGR case, this simple derivation has been chosen instead of the marginalisation of the $\chi^2$ mapping due to it being much more constraining. 
      \item $^\textrm{c}$ The first value is the direct DDGR fit, with a Gaussian uncertainty. The second one results from the $\chi^2$ mapping of solutions (Section~\ref{timing}).
    \end{tablenotes}

\end{threeparttable}
\end{adjustbox}
\end{table*}

\begin{figure*}
\centering
 \includegraphics[width=2\columnwidth]
 {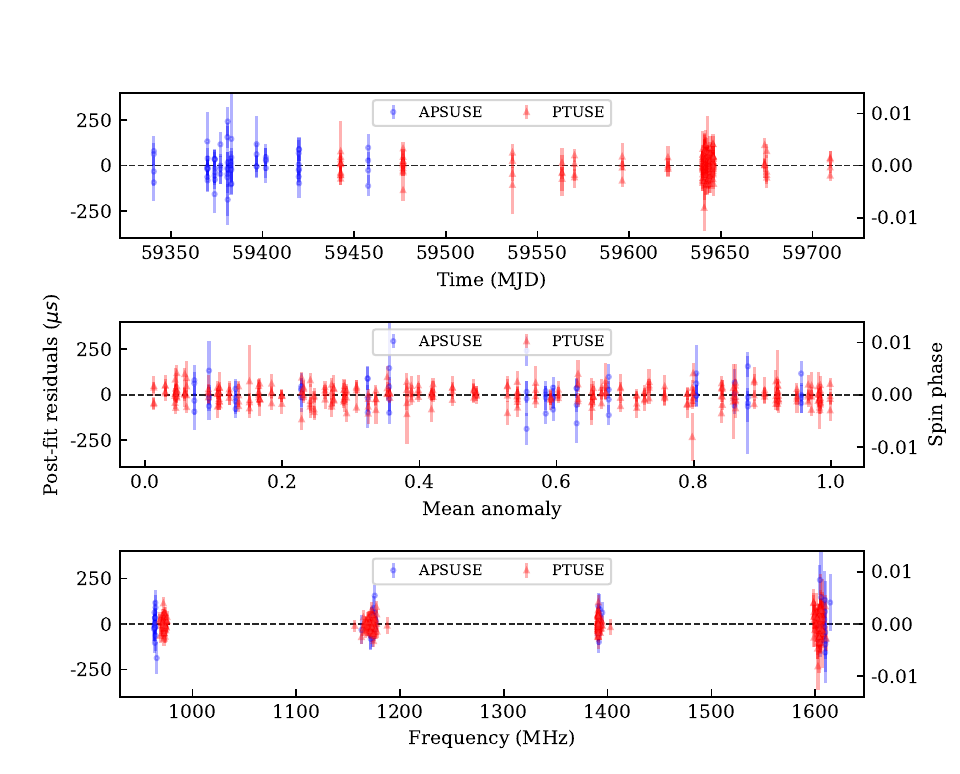}
 \caption{Timing residuals against observing time, mean anomaly, and radio frequency of the ToAs extracted from the \texttt{TempoNEST} DDH fit, presented in Table~\ref{measurements}. No significant trends are seen in the data, which is indicative of the quality of the fit. The DDGR fit presents virtually the same residuals.}
 \label{residuals}
\end{figure*}

Table~\ref{measurements} presents the timing parameters resulting from fits to the ToAs with the DDH and the DDGR models, and Fig.~\ref{residuals} shows the timing residuals. Both models are a good description of the data, with reduced $\chi^2\approx0.95$, and with spin, astrometric, Keplerian and PK parameters in very good consistency.

The results of the timing analysis are consistent with J1208$-$5936 being the first-born NS in a DNS system. It is a mildly recycled pulsar ($P_0=28.71$~ms, $\dot P\lesssim4\times10^{-20}$~s/s, Fig.~\ref{p-pdot}) in an eccentric, compact orbit ($e=0.3480$, $P_\textrm{b}=0.6316$~days) in a binary with both masses being consistent with those of NSs ($M_\textrm{t}=2.586(6)$~M$_\odot$, $M_\textrm{p}=1.26^{+0.13}_{-0.25}$~M$_\odot$, $M_\textrm{c}=1.32^{+0.25}_{-0.13}$~M$_\odot$). Similar to most massive Galactic pulsar binaries, the system lies close to the Galactic plane, at a galactic latitude of $b=2.813$~degrees. With $\textrm{DM}\approx344.4$~cm$^{-3}$pc, and assuming the NE2001\footnote{\url{https://pypi.org/project/pyne2001/}} \citep{cordes2002ne2001} or the YMW16\footnote{\url{http://119.78.162.254/dmodel/index.php}} \citep{yao2017new} models of Galactic electron density, the corresponding DM-inferred distances from Earth are $d\approx8.2(1.6)$~kpc or $d\approx8.5(1.7)$~kpc respectively and with 20\% uncertainties, placing it as possibly the furthest known Galactic DNS system.

Due to the low significance of the spin frequency derivative $\dot\nu$, both the characteristic age $\tau_\textrm{c}$ and surface magnetic field $B_\textrm{surf}$ of J1208$-$5936 are poorly constrained. Furthermore, the observed magnitude of $\dot\nu$ is likely dominated by contributions from the Shklovskii effect and the Galactic acceleration field (Section~\ref{biases_section}, Table~\ref{biases}). However, the current constraint points towards J1208$-$5936 possibly having the lowest spin-down rate among all DNS systems, indicating a weak magnetic field.

With the constrained masses and orbital parameters, we numerically integrate equations 5.6 and 5.7 in \cite{peters1964gravitational} to compute a merger time of $\tau_\textrm{m}=7.2\pm0.2$~Gyr, where the uncertainty arises from the ranges of the individual $M_\textrm{p}$ and $M_\textrm{c}$ values. Therefore, J1208$-$5936 joins the family of DNS systems merging within the Hubble time due to the orbital decay because of gravitational wave radiation (Table~\ref{DNSsystems}).

\begin{figure}
\centering
 \includegraphics[width=\columnwidth]{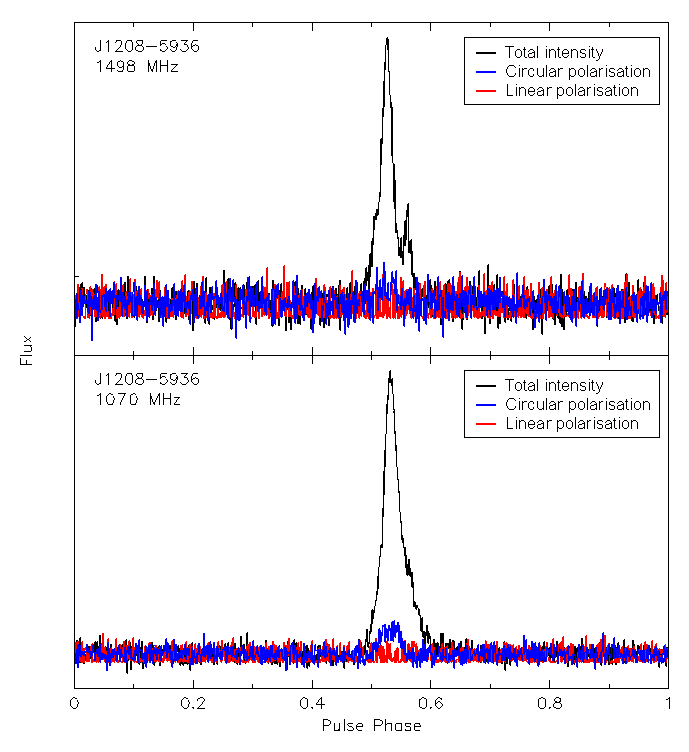}
 \caption{\label{components}De-dispersed ($\textrm{DM}=344.258$~pc\,cm$^{-3}$) PTUSE profiles at the upper and lower half of the bands of the emission of J1208$-$5936, with full Stokes resolution at $\textrm{RM}=0$~rad\,m$^{-2}$, derived with \texttt{psrchive/psrplot}. No linearly polarised emission has been detected.}
\end{figure}

\begin{figure}
\centering
 \includegraphics[width=\columnwidth]{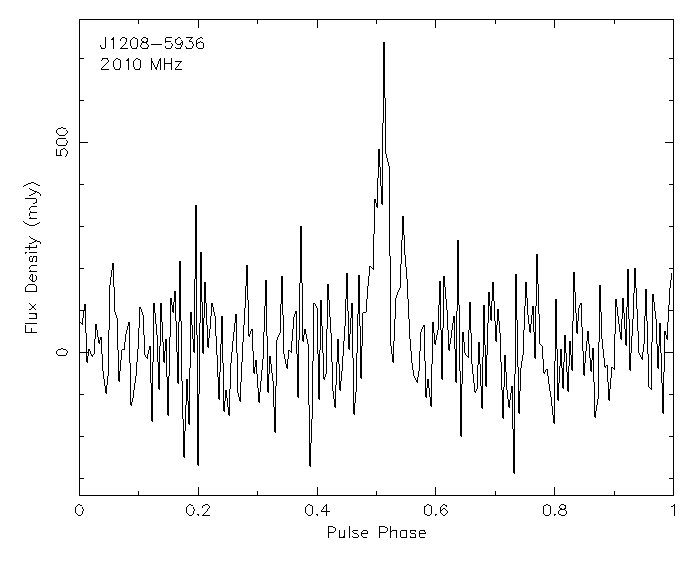}
 \caption{\label{uwl_components}De-dispersed ($\textrm{DM}=344.308$~pc\,cm$^{-3}$) UWL profile from scrunching the emission of J1208$-$5936 between 1715 and 2305 MHz. Only total intensity data is available.}
\end{figure}

\begin{figure}
\centering
 \includegraphics[width=\columnwidth]{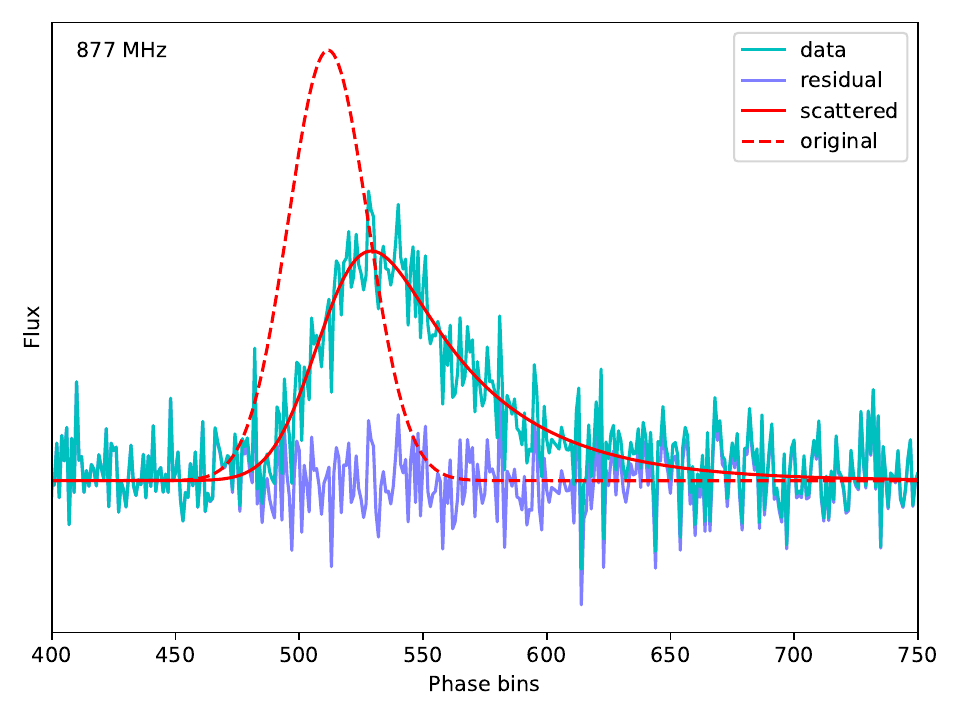}
 \caption{\label{scattered_profile}Fit of equation \ref{profile_model} (red line) at the bottom-most sub-band of 877~MHz (cyan line). The fit is effective at getting the general shape of the pulse profile,  with little structure left in the residuals (dark blue line). Table~\ref{scatter_table} lists the best scattering timescales and Gaussian widths for all fitted frequencies.}
\end{figure}

\begin{table}[h]
\caption[]{\label{scatter_table} Gaussian width $\Delta b$ and scattering timescale $\tau_\textrm{ms}$, with both units of bins (out of 1024) and ms. Aside from a couple of outliers, there is a clear increase of $\tau_\textrm{s}$ with decreasing $f$.}
\centering
\begin{tabular}{cccc}
\hline
\hline
$f$ & $\Delta b$ & $\tau_\textrm{s}$ & $\tau_\textrm{s}$ \\ 
(MHz) & (bins) & (bins) & (ms) \\
\hline
1110 & $11.0\pm0.7$ & $19.4\pm1.5$ & $0.54\pm0.04$ \\
1083 & $10.0\pm0.6$ & $20.8\pm1.3$ & $0.58\pm0.04$ \\
1057 & $11.2\pm0.6$ & $22.0\pm1.4$ & $0.62\pm0.04$ \\
1030 & $12.5\pm0.6$ & $21.4\pm1.4$ & $0.60\pm0.04$ \\
1003 & $11.3\pm0.6$ & $25.4\pm1.3$ & $0.71\pm0.04$ \\
 976 & $14.4\pm0.6$ & $21.6\pm1.4$ & $0.60\pm0.04$ \\
 955 &  $9.7\pm0.8$ & $28.1\pm2.0$ & $0.79\pm0.05$ \\
 922 & $14.3\pm0.7$ & $31.0\pm1.5$ & $0.87\pm0.04$ \\
 896 & $14.9\pm0.7$ & $33.7\pm1.6$ & $0.95\pm0.05$ \\
 877 & $16.0\pm1.2$ & $39.7\pm2.9$ & $1.11\pm0.08$ \\

\hline
\hline
\end{tabular}

\end{table}

\subsection{Profile properties}\label{profile_section}

The emission of J1208$-$5936 exhibits no detectable linear polarisation. Considering possible Faraday smearing, we searched the Rotation Measure (RM) space in the range of $-20000<\textrm{RM}<20000$~rad\,m$^{-2}$ using a 1024-channel frequency-resolved profile from the PTUSE data of the semi-coherent orbital campaign. We used both \texttt{psrchive/rmfit} and \texttt{RMcalc}\footnote{\url{https://gitlab.mpifr-bonn.mpg.de/nporayko/RMcalc}} but found no detection, which may be caused by the scattering of the pulse at low frequencies.

As seen in Fig.~\ref{components}, a secondary component trails the brighter, primary component of the pulse at high frequencies with a relative separation between peaks of approximately 12.7~deg. This secondary component disappears in the lower half of the MeerKAT band, being absorbed into the scattered tail of the main component, but becomes more prominent at high frequencies ($>$1700~MHz).
In addition to the MeerKAT observations, we also have eight fold-mode Parkes/Murriyang\footnote{\url{https://www.google.com/search?channel=fs&client=ubuntu&q=Parkes+telescope}} observations totaling more than 15 hours. They were recorded with the Ultra-Wideband (UWL) receiver, which covers the band between 704 MHz and 4032 MHz \citep{george2020uwl}. We calibrated the bandpass with the \texttt{psrchive/pac} command and the standard candle observations provided by CSIRO,\footnote{\url{https://www.parkes.atnf.csiro.au/observing/Calibration_and_Data_Processing_Files.html}} cleaned them from RFI with \texttt{clfd}, and then took the pulse profile between 1715 and 2305 MHz by adding all the observations. Fig.~\ref{uwl_components} shows the resulting pulse profile, where the secondary component is seen to gain prominence.

We find good evidence for scattering in the low-frequency band of MeerKAT observations ($<$1000~Mhz). We divide the MeerKAT band into 32 sub-bands and fit the profile in the 10 lowest frequency sub-bands with a Gaussian function convolved with an exponential
\begin{equation}\label{profile_model}S_b=\int A\times\exp\left(\frac{\left(b'-b_0\right)^2}{2\times\Delta b^2}\right)\times\exp\left(-\frac{b-b'}{\tau_\textrm{s}}\right)\textrm{d}b'
\textrm{,}\end{equation}
where $b$ stands for bins, $b_0$ for the Gaussian centre, $\Delta b$ for the standard deviation thereof, and $\tau_\textrm{s}$ for the scattering time-scale. Table~\ref{scatter_table} shows the $\tau_\textrm{s}$ value for each frequency, and Fig.~\ref{scattered_profile} shows the fit at $f=877$~MHz as an example, where it is clear that the pulse shape is well described by a scattered Gaussian function. This results in a significant trend of $\tau_{s}$ decreasing in frequency $f$ in the shape of the power law
\begin{equation}\tau_\textrm{s}=693\pm12\textrm{ \textmu s }\left(f/\textrm{GHz}\right)^{-2.8\pm0.2}\textrm{.}\end{equation}
The scattering index $2.8\pm0.2$ is lower than the $4.0$ typically measured in single-component scattered pulsars \citep[e.g.][]{oswald2021scatter}, which is likely indicative of a bias introduced by the double-component nature of J1208-5936. Nonetheless, our analysis provides robust evidence in favour of the presence of scattering in the pulse. Finally, from the PTUSE 1024-channel profile, the best DM measurement from the 2 March 202 to 8 March 2022 observational campaign is $344.298\pm0.054$~pc\,cm$^{-3}$ from the \texttt{pdmp} fit at full frequency resolution.

\subsection{Comparison with the known DNS population}

\begin{figure}
\centering
 \includegraphics[width=\columnwidth]{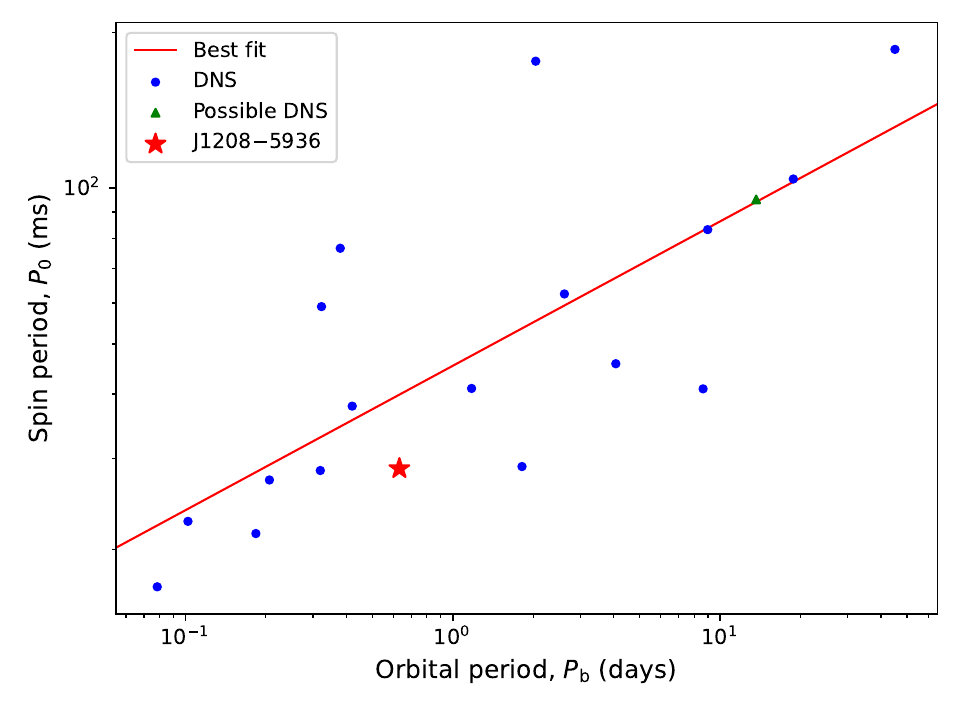}
 \caption{$P_0$ - $P_\textrm{b}$ diagram of the known recycled pulsars in DNS systems population. The lines represent linear regression fits to the data points in the log space, added to aid in the visualisation of the trend.}
 \label{p-pb}
\end{figure}

\begin{figure}
\centering
 \includegraphics[width=\columnwidth]{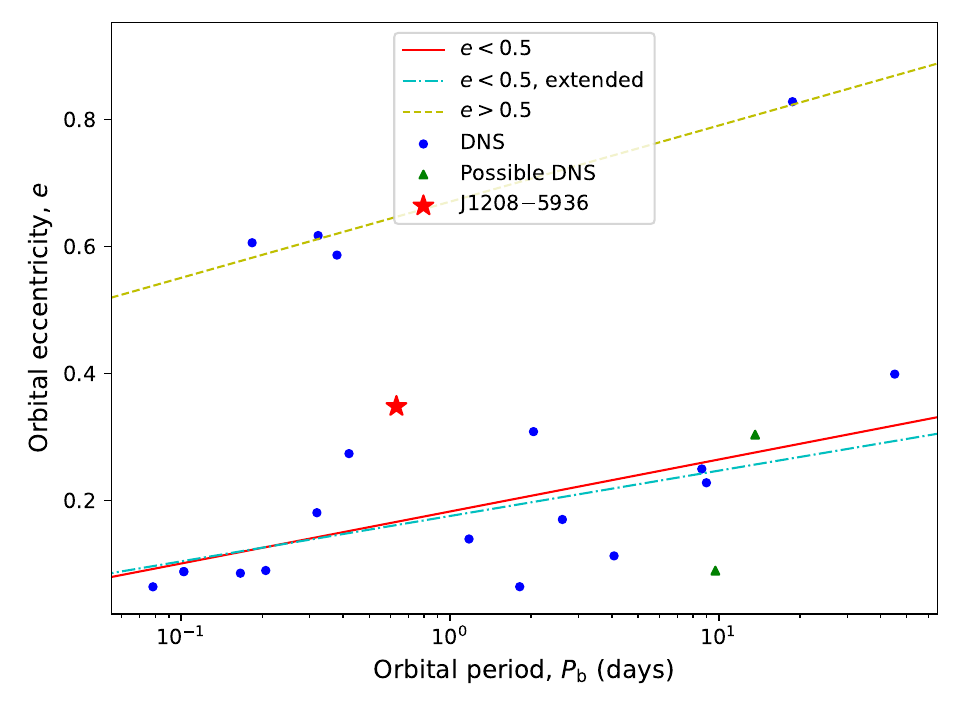}
 \caption{$e$ - $P_\textrm{b}$ diagram of the known pulsars in DNS systems population. The lines represent linear fits to the data points at $e<0.5$ and $e>0.5$, with the extended fit including unconfirmed DNS systems.
 }
 \label{e-pb}
\end{figure}

J1208$-$5936 falls on the lower side of the expected $P_0$ and $P_\textrm{b}$ relationship, having a spin period below the average (Fig.~\ref{p-pb}), but it is still consistent with the rest of the DNS population. This relationship is explained through the less efficient recycling of the primary NS resulting from a delayed RLO accretion onset in longer-orbit progenitor NS - He star systems \citep{tauris2017formation}.

A relationship between $e$ and $P_\textrm{b}$ has been postulated on a similar basis \citep{tauris2015ultra,tauris2017formation}. In this case, wide NS - He star progenitors undergo a reduced mass transfer due to the delayed RLO onset, which results in a less stripped He star and therefore in an increased mass-loss during the second supernova. Therefore, large orbital periods are associated with large eccentricities. Fig.~\ref{e-pb} shows that J1208$-$5936 is a high-eccentricity outlier amongst DNS systems with $e<0.5$ with short orbital periods. Such outliers are to be expected due to the large spread in outcome eccentricities introduced by supernova kicks \citep{tauris2017formation}, but it could also be a hint towards larger supernova kick (see Section~\ref{formation}).

\begin{figure}
\centering
 \includegraphics[width=\columnwidth]{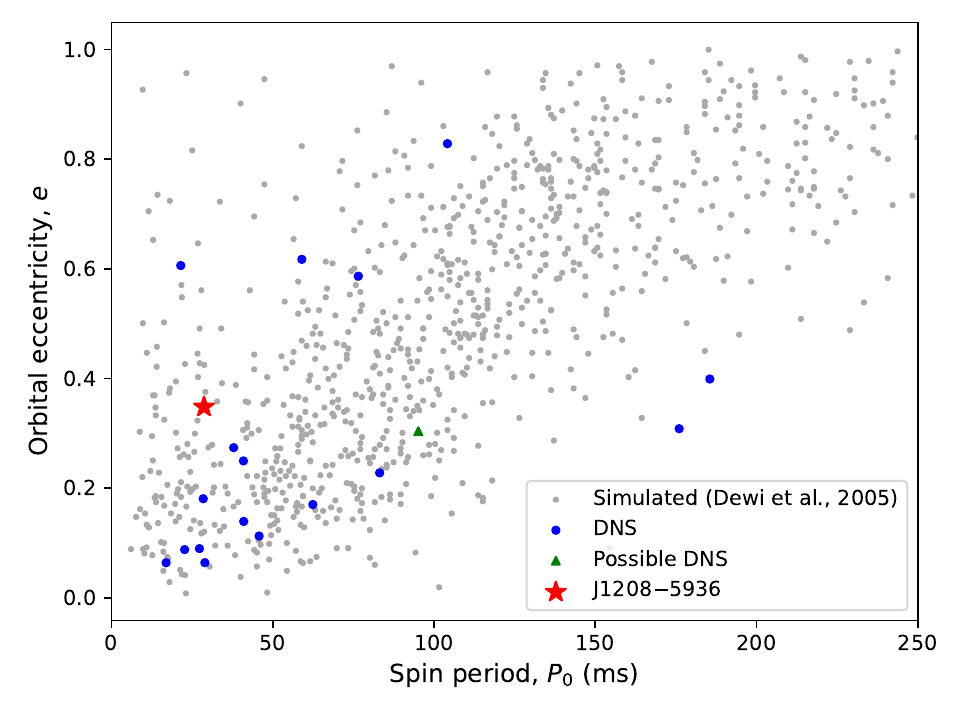}
 \caption{$e$ - $P_0$ diagram of the known recycled pulsars in DNS systems population. The spread of values in observed systems is larger than predicted in some simulations \citep{dewi2005spin}, with J1208$-$5936 staying slightly above the expectation.}
 \label{e-p}
\end{figure}

Finally, we also look at the postulated $e-P_0$ relationship in Galactic DNS systems. This one goes in parallel with the $e-P_\textrm{b}$ relationship, and is explained from the $P_0-P_\textrm{b}$ relationship: pulsars in longer orbital periods undergo less efficient recycling, the He star undergoes greater mass loss during the supernova, and therefore high eccentricities should imply longer spin periods. This was first observed by \cite{faulkner2004psr}, and then theorised by \cite{dewi2005spin} under the assumption of supernova kicks smaller than 50 km\,s$^{-1}$. Fig.~\ref{e-p} compares the updated DNS population with the simulated results from \cite{dewi2005spin}, and it shows that J1208$-$5936 is once again at the high-eccentricity end but still lies within the distribution. As recently reported by \cite{sengar2022high}, the simulations from \cite{dewi2005spin} do not coincide with the observed high-eccentricity end of the Galactic DNS population, which would favour the different formation channel hypothesis brought forward in \cite{andrews2019double}.

\begin{figure}
\centering
 \includegraphics[width=\columnwidth]{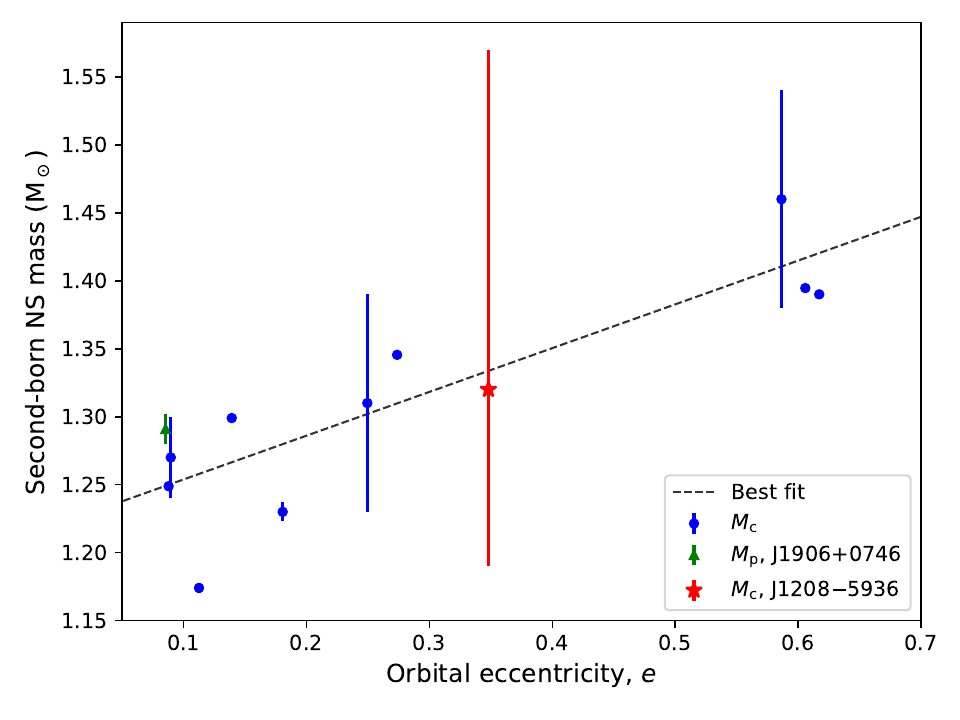}
 \caption{$M_\textrm{c}$ - $e$ diagram of second-born NSs with measured masses (quoting the mass derived from the $\chi^2$ mapping of DDGR solutions for J1208$-$5936). The line represents linear fits to the data points excluding the companion of J1208$-$5936, which does not yet have a sufficiently constrained mass. For PSR J1906$-$0746 we quote the pulsar mass instead of the companion mass because it is believed to be the second-born, un-recycled NS.
 }
 \label{m2-e}
\end{figure}

\subsection{Formation channel}\label{formation}

We note in Fig.~\ref{e-pb} that J1208$-$5936 has a particularly high orbital eccentricity compared to other DNS systems with $e<0.5$. This could be indicative that J1208$-$5936 has been formed through a different channel than other systems in this group. We also pay attention to the division between $e>0.5$ and $e<0.5$ systems in the Galactic DNS populations in the $e-P_\textrm{b}$ space (Fig.~\ref{e-pb}). According to \cite{andrews2019double}, the high-eccentricity population can be explained by larger supernova kicks from heavier He stars progenitors of the second-born NS. \cite{tauris2015ultra} also explores the possible evolutionary origin of this division. In most known DNS systems, the He star loses enough mass during binary interaction to stop the possibility of nuclear fusion in the ONeMg core. In these cases, the He star with a core mass $\lesssim$$1.43$~M$_\odot$ would implode through the electron-capture process instead of following shell burning until reaching the iron core collapse \citep{tauris2015ultra}. On the other hand, He stars that keep a core with $\gtrsim$$1.43$~M$_\odot$ would be able to reach iron core collapse, possibly leaving a heavier NS behind with a larger supernova kick and with greater mass loss \citep{tauris2015ultra}. In this picture, most known DNS systems that do not get disrupted form through the electron-capture channel, constituting the $e<0.5$ population, while the DNS systems forming through the iron core collapse channel could create a heavier $e>0.5$ population. This picture is in principle corroborated by Fig.~\ref{m2-e}, where the high eccentricity systems are also shown to contain more massive second-born NSs, consistent with the idea of larger supernova kicks being associated with larger masses.

It is tempting to see J1208$-$5936 as a system bridging the gap between these two postulated populations. While its current eccentricity is consistent with the tail of the eccentricity distribution with a second supernova kick of 50~km\,s$^{-1}$ and a progenitor He stars mass of 3~M$_\odot$ like other low-eccentricity DNS systems \citep[see Fig.~10 in][]{tauris2017formation}, it is also plausible that its companion has formed through the iron core collapse channel, or that it has at least suffered a stronger supernova quick than in other $e<0.5$ DNS systems formed through the electron capture process owing to a more massive He star progenitor. However, in Fig.~\ref{m2-e} it is shown that the current uncertainties on $M_\textrm{c}$ are too large to discriminate whether J1208$-$5936 lies within any of these two populations. Further timing in the following years will constrain the mass of the companion of J1208$-$5936, providing clarification on its formation channel.

\section{Search for pulsations from the companion}\label{companion_search}

We also searched for pulsations from the companion of J1208$-$5936. Detecting them would not only increase the sample of known young pulsars in DNS systems, but it would also provide much more precise mass measurements and even gravity tests in the future through the inclusion of the mass ratio as an extra constraint on top of the PK parameters, such as in the case of PSR J0737$-$3039A/B \citep{lyne2004double}.

We performed deep searches on de-modulated APSUSE data from 30-minute and 60-minute long observations from the Shapiro delay semi-coherent orbital campaign. We began by cleaning the data with \texttt{presto/rfifind}, de-dispersing it and integrating it into a barycentric time series at $\textrm{DM}=344.43$~pc~cm$^{-3}$ with \texttt{presto/prepdata}. This DM value is slightly offset from our current best estimate presented in Section~\ref{profile_section}, owing to the fact that we are dealing with APSUSE data instead of PTUSE. Nonetheless, small discrepancies in DM do not affect the search for pulsations at $P>10$~ms. We de-modulated the time series with \texttt{python} software \texttt{pysolator}\footnote{\url{https://github.com/alex88ridolfi/pysolator}}, which undoes the orbital motion of the companion assuming a chosen mass ratio, allowing us to search for pulsations in its rest frame of the companion. For this, we assumed several system mass ratios chosen from 27 different inclination angles between $i=45^\circ$ and $i=85^\circ$ uniformly spaced in $\sin{i}$, assuming the total mass of $M_\textrm{t}=2.586$~M$_\odot$ from the measurement of $\dot\omega$ (Section~\ref{timing}, Table~\ref{measurements}). The spacing in $\sin{i}$ was chosen on the basis of introducing a maximum line-of sight accelerations discrepancy of at most 1~m\,s$^{-2}$ between the different de-modulations based on different $\sin{i}$ values.

Subsequently, two different searches were performed on this data set. Firstly, we computed the FFT of each de-modulated time series with \texttt{presto/realfft} and then de-reddened them with \texttt{presto/rednoise}. The de-reddened FFT spectra were searched with \texttt{presto/accelsearch} with $z_\textrm{max}=2$. For a 30-minute-long and a 60-minute-long observations, this corresponds to acceleration ranges of $\pm$1.85~m\,s$^{2}$ and $\pm$0.46~m\,s$^{2}$ for 10-ms pulsar, with increasing ranges for increasing spin period. All candidates were then refolded with \texttt{presto/prepfold}.

The second search implemented a fast Folding algorithm (FFA) instead, a phase-coherent periodicity search technique on evenly sampled time series \citep{staelin1969fast}. The FFA has the advantage of being more sensitive to long-period, narrow signals than the FFT \citep[e.g.][]{cameron2017investigation}. Additionally, since it does not compute a Fourier Transform, de-reddening is typically performed with a running-median filter on the time series, which is less harmful towards real low-frequency signals than removing power from Fourier bins \citep{singh2022gmrt}. Therefore, the FFA is well-suited for searching for a non-recycled second-born pulsar in the properly de-modulated time series. We restricted our FFA search to periods above 100 ms, on the basis that shorter periods would have easily been detected by the \textsc{presto} FFT search. We used the \textsc{riptide}\footnote{\url{https://github.com/v-morello/riptide}} software \citep{morello2019high} to perform a non-accelerated FFA search on the de-modulated time series and fold the resulting candidates. For that we implemented the following steps with our \texttt{presto} wrapper pipelines \texttt{demodulate-search}\footnote{\url{https://github.com/mcbernadich/demodulate-search}}.

No significant pulsations from the companion were found. This suggests that the companion has either crossed its pulsar death line, that its radio emission is beamed away from Earth, or that the pulses are too faint to be detected with the available data and telescope sensitivity.

\section{Implications for NS merger rate}\label{merger_implications}

The discovery of J1208$-$5936 and, more significantly, the increase in explored depth of the southern Galactic plane provided by the sensitivity of the MMGPS-L survey are a great opportunity to update the prediction of the observed NS coalescence events with gravitational wave observatories such as LIGO and Virgo. This is possible assuming that the observed population of pulsars in DNS systems is representative of the broad Galactic DNS population, with each new survey and discovery providing a more accurate sampling of it.

Following the methodology established in \cite{kim2003probability} and also implemented in \cite{kim2010effect,kim2015implications,pol2019future,pol2020updated,grunthal2021revisiting}, we use \texttt{PsrPopPy2}\footnote{\url{https://github.com/devanshkv/PsrPopPy2}} to populate the Galactic field with DNS systems and simulate blind surveys on the sky. In this case, we simulate the known DNS systems merging within the Hubble time. For each known DNS system $j$, we seek to find the proportionality constant $\alpha_j$ in
\begin{equation}\label{rates}\lambda_j=\alpha_j N_{\textrm{tot,}j}\textrm{,}\end{equation}
where $\lambda_j$ is the number of discovered systems and $N_{\textrm{tot,}j}$ is the number of simulated systems with similar characteristics each. This $\alpha_j$ value is then used to compute a probability distribution for the merger rate $\pazocal{R}_j$
\begin{equation}\label{R_distribution}P\left(\pazocal{R}_j\right)=\left(\frac{\alpha_j\, \tau_{\textrm{life},j}}{f_{\textrm{b},j}}\right)^2 \pazocal{R}_j\times\exp{\left(-\frac{\alpha_j\,\tau_{\textrm{life},j}}{f_{\textrm{b},j}}\pazocal{R}_j\right)}\textrm{,}\end{equation}
where $\tau_{\textrm{life},j}=\tau_\textrm{age}+\tau_\textrm{obs}$ is the lifetime of the system, composed by its age and future observable time, and $f_\textrm{b}$ is beaming factor of the pulsar (inverse of the beam sweep fraction). Typically, $\tau_\textrm{age}$ is defined either by the characteristic age ($\tau_c$) of the pulsar or the time after it has exited a recycling fiducial line, while $\tau_\textrm{obs}$ consists either on the remaining time until the pulsar crosses the death-line or the merger time ($\tau_\textrm{m}$).
Finally, the rate of NS merger events in the Milky Way ($\pazocal{R}_\textrm{MW}$) is then the sum of all the $\pazocal{R}_j$ computed for all DNS systems merging within the Hubble time, which for the probability distribution translates to the convolution of all probability distributions.

The latest update on the NS merger rate prediction \citep{grunthal2021revisiting} includes the following blind pulsar surveys: the Pulsar Arecibo L-band Feed Array survey \citep[PALFA,][]{cordes2006arecibo}, the Low-latitude High Time-Resolution Universe pulsar survey \citep[HTRU-Low,][]{keith2010high}, the Parkes High-latitude pulsar survey \citep{burgay2006parkes}, the Parkes Multibeam Survey \citep{manchester2001parkes}, the PSR B1534+12 discovery survey with Arecibo \citep{wolszczan1991nearby} and the Green
Bank North Celestial Cap survey \cite[GBNCC,][]{stovall2014green}. These surveys used to provide a realistic coverage of the explored pulsar sky, and they all have in common that they were performed with single-dish telescopes and therefore single-beam pointings on the sky, which are straightforward to model in \texttt{PsrPopPy2} as circular beams with a Gaussian sensitivity function. Now we update this picture with the MMGPS-L, the most sensitive survey in the southern Galactic plane. Its interferometric nature requires an extra layer of complexity in the simulation of pointings in the sky, which we implement in \texttt{PsrPopPy2} with the addition of an extra degradation factor for coherent beams (see Appendix~\ref{beamPatternPsrPopPy2} for details).

The biggest uncertainty in accounting for J1208$-$5936 in the simulations is the computation of $\tau_\textrm{life,J1208}$. Thanks to the preliminary mass estimates presented in Section~\ref{timing}, we compute a merger time of $\tau_\textrm{m}=7.2\pm0.2$~Gyr for J1208$-$5936. However, unlike with the rest of pulsars merging within the Hubble time, there is little information on the characteristic age $\tau_\textrm{c}$ of J1208$-$5936 due to the poorly constrained $\dot P$. Despite this, a reasonable assumption for $\tau_\textrm{age}$ can be made without sacrificing accuracy. Due to its large merger time, J1208$-$5936 is already the pulsar with the smallest contribution to the estimated galactic merger rate, and therefore we do not expect the choice to have a significant impact when all contributions are added. Given its small $\dot P$, it is likely a pulsar comparable to PSR J1913+1102 (Table~\ref{DNSsystems}). We therefore pick a realistic recycling age of 2.5~Gyr. For the beaming factor, we choose $f_\textrm{b,J1208}=4.59$, consistent with the average of pulsars in DNS systems with actually measured $f_{\textrm{b},i}$ values \citep{pol2019future}, and model the pulse duty fraction of $\delta_\textrm{J1208}=8$\% from the part of the profile that was visible above the noise during the discovery. Finally, we compute its Doppler degeneration factor $0\leq \gamma_{2m}\leq1$ in all surveys, assuming acceleration search for each of them (see Appendix~\ref{beamPatternPsrPopPy2} for more details).

\begin{table}
\begin{threeparttable}
\caption[]{\label{NSmerger_table} Pulsars used in the NS merger rate computation simulation, along with their used lifetime, beaming fraction, pulse duty ($\delta$, used only during the \texttt{PsrPopPy2} runs), and the derived $\alpha_i$ and individual merger rate contributions. For PSR J1906+0746, the lifetime is shorter than the merger time due to it crossing the pulsar deadline in $\sim$60~Myr.}
\centering
\begin{tabular}{lccccc}
\hline
\hline
PSR$_j$ & $\tau_{\textrm{life,}j}$ & $f_{\textrm{b},j}$ & $\delta_j$ & $\alpha_j$ & $\pazocal{R}_j$ \\ 
 & (Myr) &  & (\%) & & (Myr$^{-1}$) \\
\hline
J1946+2052    & 293    & 4.59\,$^\textrm{c}$ & 6    & 0.00609 & 2.6$^{+11.7}_{-1.88}$ \\[.4ex]
J1757$-$1854  & 235    & 4.59\,$^\textrm{c}$ & 6    & 0.00916 & 2.15$^{+9.67}_{-1.49}$ \\[.4ex]
J0737$-$3039A & 244    & 2.0\,$^\textrm{a}$  & 27   & 0.00429 & 1.92$^{+8.7}_{-1.41}$ \\[.4ex]
B1913+16      & 377    & 5.7\,$^\textrm{a}$  & 16.9 & 0.00914 & 1.66$^{+7.55}_{-1.15}$ \\[.4ex]
J1906+0746    & 60     & 1.0\,$^\textrm{b}$  & 1    & 0.00368 & 4.49$^{+20.71}_{-3.26}$ \\[.4ex]
J1913+1102    & 3\,125 & 4.59\,$^\textrm{c}$ & 6    & 0.01010 & 0.14$^{+0.67}_{-0.09}$ \\[.4ex]
J0509+3801    & 729    & 4.59\,$^\textrm{c}$ & 18   & 0.01008 & 0.62$^{+2.86}_{-0.83}$ \\[.4ex]
J1756$-$2251  & 2\,086 & 4.59\,$^\textrm{c}$ & 3    & 0.01290 & 0.17$^{+0.77}_{-0.11}$ \\[.4ex]
B1534+12      & 2\,908 & 6.0\,$^\textrm{a}$  & 4    & 0.01415 & 0.14$^{+0.67}_{-0.09}$ \\[.4ex]
J1208$-$5936  & 9\,700 & 4.59\,$^\textrm{c}$ & 8    & 0.01157 & 0.04$^{+0.19}_{-0.03}$ \\

\hline
\hline
\end{tabular}

\begin{tablenotes}
      \small
      \item $^\textrm{a}$ Beaming fractions obtained from Table 2 in \cite{kim2015implications}.
      \item $^\textrm{b}$ Beaming fraction set to 1.0 as we implement explicit modelling of its variation in time. See \cite{grunthal2021revisiting} for more details.
      \item $^\textrm{c}$ Beaming factor taken from the average of J0737-3039A, B1913+16 and B1534+12.
    \end{tablenotes}

\end{threeparttable}
\end{table}

\begin{figure}
\centering
 \hspace*{-0.3cm} 
 \includegraphics[width=1.01\columnwidth]{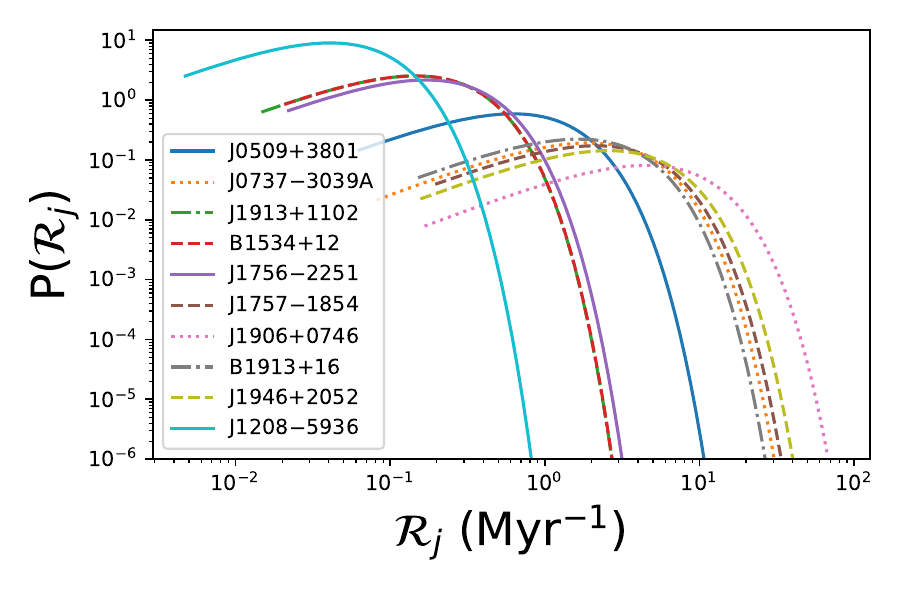}
 \includegraphics[width=\columnwidth]{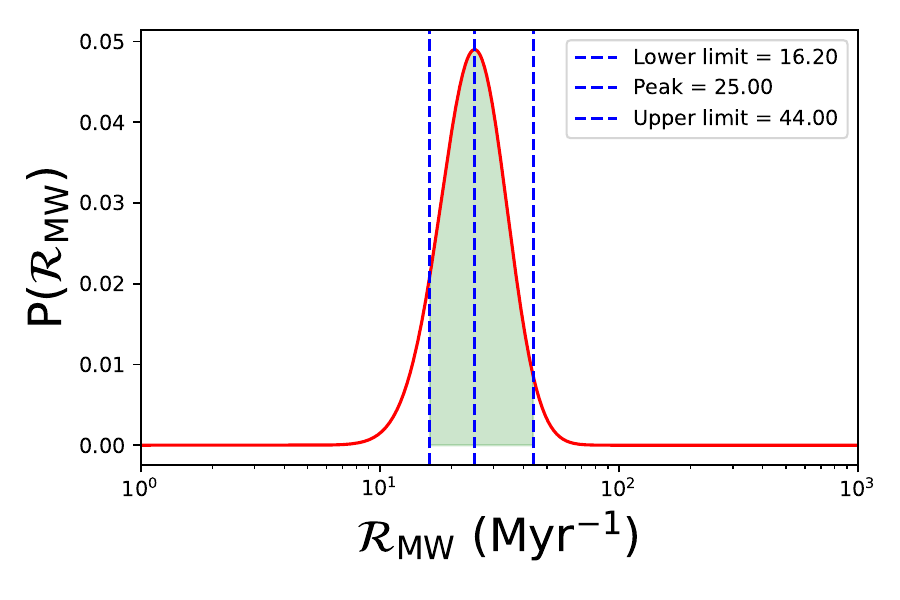}
 \caption{Updated probabilities distribution of the merger rates after the inclusion of J1208$-$5936 and the MMGPS-L survey. \textbf{Top:} individual contributions from each DNS population, derived from the values in Table~\ref{NSmerger_table} and equation~\ref{R_distribution}. The 90\% credible intervals are quoted in Table~\ref{NSmerger_table}. \textbf{Bottom:} Galactic merger rate probability distribution from the convolution of individual DNS distributions, with a highlighted 90\% credible interval.}
 \label{NSmerger_plot}
\end{figure}

For consistency with \cite{pol2019future,pol2020updated,grunthal2021revisiting}, we simulate pulsars with a mean luminosity of $\langle\log_{10}{\left(L/\left[\textrm{mJy\,kpc}^2\right]\right)}\rangle=-1.1$ with a standard deviation of $\sigma_{\log_{10}L}=0.9$, and a mean spectral index of $\langle\Gamma\rangle=1.4$ with a deviation of $\sigma_\Gamma=1$. The height scale of the simulated population is set to $z_0=0.33$ kp, with pulsars following a density distribution above and below the Galactic plane of $f\left(z\right)=\exp{\left(-z/z_0\right)}$. For PSR J1906+0746, we use the beam shape modelling implemented in Section 4 of \cite{grunthal2021revisiting}.
From each DNS system $j$, we start with simulating only $N_{\textrm{tot},j}=100$ pulsars on the sky and simulate the surveys to see the number of discoveries $\lambda_j$. Then, on each step, the number of pulsars is increased by $\Delta N_{\textrm{tot},j}=100$ and the surveys are repeated until we reach $N_{\textrm{tot},j}=4\,000$. Within each step, the population is simulated 100 different times, and in each simulation, the surveys are performed 100 different times, producing a loop with 10\,000 iterations. All of this is done to get an averaged measurement of $\alpha_j$ from equation~\ref{rates}.

The updated contributions to the Galactic merger rate, along with the parameters used in the simulations and computations, are presented in Fig.~\ref{NSmerger_plot} and Table~\ref{NSmerger_table}. PSR J1906+0746 remains the most impactful contribution due to its short lifetime, while the contribution of J1208$-$5936 is the smallest due to its large lifetime. The added Galactic merger rate results in \begin{equation}\pazocal{R}_\textrm{MW}^\textrm{new}=25^{+19}_{-9}\textrm{\,Myr}^{-1}\end{equation}
with a 90\% confidence interval. This result is shifted downwards with respect to the $\pazocal{R}_\textrm{MW}^\textrm{2021}=32^{+19}_{-9}$~Myr$^{-1}$ presented by \cite{grunthal2021revisiting}, owing to the inclusion of the MMGPS-L survey, because despite it being the most sensitive blind pulsar survey on the southern Galactic plane, it has only added one DNS system merging within Hubble time, therefore implying a reduction of the expected number of unseen merging DNS systems in the Milky Way from the lack of new detections.

\begin{figure}
\centering
 \includegraphics[width=\columnwidth]{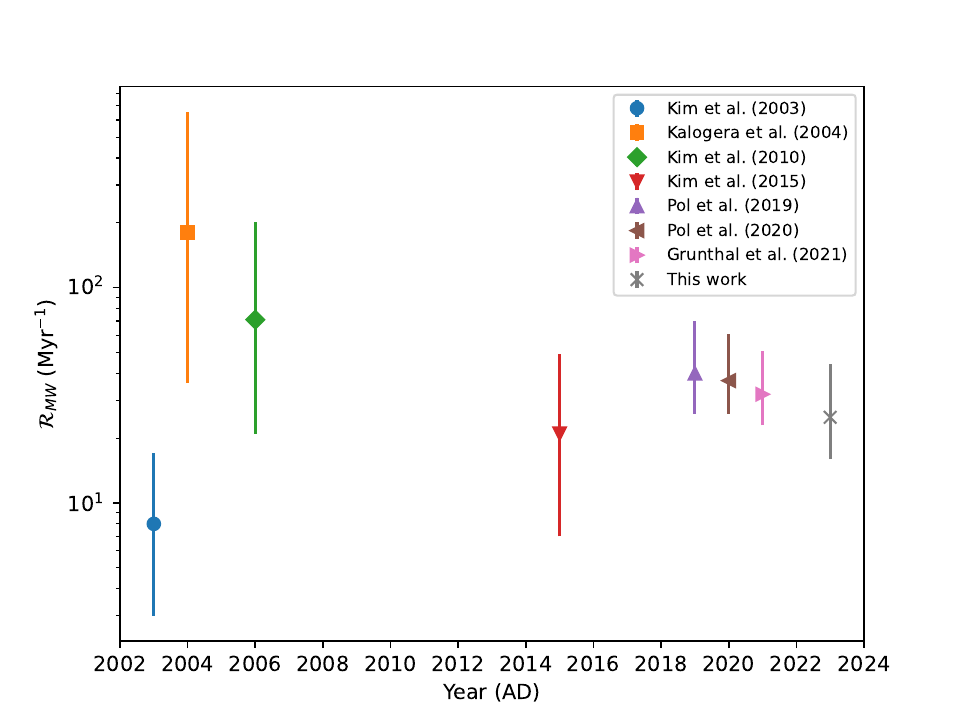}
 \caption{Evolution of estimated NS merger rates in the Milky Way based on observable Galactic binaries during the last two decades, quoting the 90\% or 95\% confidence intervals \citep{kim2003probability,kalogera2004cosmic,kim2010effect,kim2015implications,pol2019future,pol2020updated,grunthal2021revisiting}.}
 \label{NSmerger_comparisson}
\end{figure}

We pay attention to the increasingly constrained NS merger rates in the Milky Way based on the observation of Galactic binaries, as well as the decreasing trend of the estimated NS merger rates in the Milky Way based on the observation of Galactic binaries. As shown in Fig.~\ref{NSmerger_comparisson}, the discovery of PSR J0737$-$3039 with its relatively short merger time brought forward a drastic increase of the predicted Galactic merger rate \citep{burgay2006j0737,kalogera2004cosmic,kim2010effect}) with respect to previous estimates based only on PSR B1913+16 and PSR B1534+12 \citep[eg.,][]{kim2003probability}. However, further discoveries have not increased the estimated rate in recent years, as better modelling of beam shapes and sky coverage have allowed for an increase inn the accuracy of the estimates. \cite{pol2019future} presented $\pazocal{R}_\textrm{MW}^{2019}=42^{+30}_{-14}$~Myr$^{-1}$ at the 90\% confidence with the inclusion of at-the-time newly discovered systems such as of PSR J1757$-$1854 and the highly relativistic PSR J1946+2052, the most constrained estimate up to that date owing to the increased coverage of the sky. Then, in \cite{pol2020updated} the sensitivity of the newly included the GBNCC survey \citep{stovall2014green} outweighed the inclusion of the newly discovered relativistic J0509+3801 DNS system, decreasing the estimated number of unseen binaries in the sky and reducing the N merger rate estimate to $\pazocal{R}_\textrm{MW}^{2020}=37^{+24}_{-11}$~Myr$^{-1}$, with more constrained uncertainties. \cite{grunthal2021revisiting} implemented a modified beam shape in their simulations, leading to another reduction of the total value and uncertainties. Our estimate presents a continuation of this trend, being our estimate once again reduced in comparison to the most recent estimates owing to the sensitivity of the MMPGS-L, the most important newly contributing factor. Therefore, our work confirms that NS merger rate estimates based on know electromagnetic binaries are converging into more constrained, lower values as pulsar surveys scout the sky at larger depths.

We transform this rate into a local cosmic merger rate density and a prediction for the LIGO event detection rates by assuming that the number of DNS systems merging within the Hubble time in a galaxy is proportional to its total B-band luminosity, as it is a tracer of the star-formation rate \citep{kopparapu2008host}. This results in a local merger rate density of
\begin{equation}\pazocal{R}_\textrm{local}^\textrm{new}=293^{+222}_{-103}\textrm{\,Gpc}^{-3}\textrm{\,yr}^{-1}\textrm{,}\end{equation}
which results on an upper limit of $\pazocal{R}_\textrm{local}\le515$~Gpc$^{-3}$yr$^{-1}$. Assuming that LIGO has sensitivity to a range distance of $D_\textrm{r}=130$~Mpc during the O3 run\footnote{\label{ligo_plans}\url{https://observing.docs.ligo.org/plan/}} (see \citealt{chen2021distance} for a definition of $D_\textrm{r}$), we predict a LIGO NS merger detection rate of 
\begin{equation}\pazocal{R}_\textrm{LIGO,O3}^\textrm{new}=2.76^{+2.10}_{-0.97}\textrm{~yr}^{-1}\textrm{.}\end{equation}
Based on the observed events between April 1st, 2019 and October 1st, 2019 catalogued in the Gravitational-Wave Transient Catalogue-2.1, a NS merger rate density of $\pazocal{R}_\textrm{local}^\textrm{GRTC-2.1}=256^{+510}_{-237}$~Gpc$^{-3}$\,yr$^{-1}$ is computed, which for detections translates into $\pazocal{R}_\textrm{LIGO,O3}^\textrm{GRTC-2.1}=2.36^{+4.69}_{-2.18}$\,yr$^{-1}$ \citep{abbott2021gwtc}, in good consistency with our estimate. This shows that the rates computed from electromagnetic binaries are in good consistency with gravitational wave observatories (see also \citealt{pol2020updated,grunthal2021revisiting}). With this in mind, we assume $D_\textrm{r}=175$~Mpc to make a prediction of the rate NS merger events seen by the LIGO-Virgo-KAGRA O4 run,$^\textrm{\ref{ligo_plans}}$ resulting in
\begin{equation}\pazocal{R}_\textrm{LIGO,O4}^\textrm{new}=6.73^{+5.12}_{-2.37}\textrm{~yr}^{-1}\textrm{,}\end{equation}
which implies a prediction of the detection of $10^{+8}_{-4}$ events during the 18 months of the O4 run,$^\textrm{\ref{ligo_plans}}$ or at most 18 events within 90\% credible intervals.

\section{Biases and future prospects}\label{biases&prospects}

\subsection{Proper motion and galactic field}\label{biases_section}

\begin{table}
\caption[]{\label{biases} Possible biases onto spin a PK parameters measurements introduced by the Shklovskii effect and the acceleration in the Galactic field (eq.~\ref{shk_acc}) and from proper motion (eq.~\ref{omdotPM}~and~\ref{xdotPM}), assuming $d=8.2$~kpc and $|\mu_\textrm{t}|=6$~mas\,s$^{-1}$ (see text in Section~\ref{biases}).}
\centering
\begin{tabular}{lccc}
\hline
\hline\\[-1.5ex]
Parameter & Shklovskii & Galactic & Proper motion \\[0.25ex]
\hline\\[-1.5ex]
$\dot P$~(s\,s$^{-1}$) & $2\times10^{-20}$ & $-2\times10^{-20}$ & ... \\
$\dot P_\textrm{b}$~(s\,s$^{-1}$) & $4\times10^{-14}$ & $-4\times10^{-14}$ & ...  \\
$\dot x$~(ls\,s$^{-1}$) & $3\times10^{-18}$ & $-3\times10^{-18}$ & $3\times10^{-15}$  \\
$\dot \omega$~(deg\,yr$^{-1}$) & ... & ... & $2\times10^{-6}$  \\[0.25ex]
\hline
\hline
\end{tabular}
\end{table}

There are three main contributions that could be biasing our mass constraints. Firstly, we consider the Shklovskii and the Galactic acceleration field effects \citep{shklovskii1970possible,damour2992orbital}. These two combine into an apparent evolution of any periodicity or length $L=\{P_0,P_\textrm{b},x\}$, expressed with the equation
\begin{equation}\label{shk_acc}\frac{\dot L}{L}=-\left(\frac{\dot D}{D}\right)_\textrm{Shkl}-\left(\frac{\dot D}{D}\right)_\textrm{Gal}=\frac{1}{c}\left(\frac{V^2_\textrm{t}}{d}+\vec{K_0}(\vec{a}_\textrm{PSR}-\vec{a}_\textrm{SSB})\right)\textrm{,}\end{equation}
where $D$ is the Doppler factor, $\vec{K_0}$ is the unit vector from the Solar System barycentre (SSB) to the pulsar system, $\vec{a}_\textrm{PSR}$ and $\vec{a}_\textrm{SSB}$ are the Galactic acceleration vectors at the pulsar system and at the SSB, $V_\textrm{t}$ is the magnitude of the transverse velocity of the system with respect to the Sun, and $d$ is the distance to the source. Secondly, we also consider the effects on $\dot\omega$ and $\dot x$ introduced by the proper motion (PM) vector $\vec{\mu_\textrm{t}}=(\mu_\textrm{RA},\mu_\textrm{DEC})$, along with the angle of the ascending node $\Omega_\textrm{a}$ and inclination angle $i$, expressed in \cite{kopeikin1996proper} as
\begin{equation}\label{omdotPM}\dot\omega^\textrm{PM}=2.78\times10^{-7}\csc{ i}\left(\mu_\textrm{RA}\cos{\Omega_\textrm{a}}+\mu_\textrm{DEC}\sin{\Omega_\textrm{a}}\right)\end{equation}
and
\begin{equation}\label{xdotPM}\left(\frac{\dot x}{x}\right)^\textrm{PM}=1.54\times10^{-16}\cot{i}\left(-\mu_\textrm{RA}\sin{\Omega_\textrm{a}}+\mu_\textrm{DEC}\cos{\Omega_\textrm{a}}\right)\textrm{\,.}\end{equation}

In our case, since we do not have a measurement of the proper motion or $\Omega_\textrm{a}$, these effects are not yet quantifiable, but we can nonetheless estimate the likely magnitude of the contributions. From the Galactic coordinates and DM distance of J1208$-$5936, we take the Galactic rotation velocity curve from \cite{sofue2020rotation} and predict $V_\textrm{t}\simeq240$~km\,s$^{-1}$ and $|\mu_\textrm{t}|\simeq6$~mas\,s$^{-1}$. This assumption is taken based on the small magnitude of the supernova kicks introduced during the formation of the system, as otherwise the binary would have been disrupted \citep{tauris2017formation}, making the Galactic flow the more dominant component of the acceleration. For the Galactic acceleration field, we use the Galactic mass distribution model presented in \cite{mcmillan2017mass} and extract $\vec{a}_\textrm{PSR}$ and $\vec{a}_\textrm{SSB}$.

Our estimation of the magnitude of these contributions are listed in Table~\ref{biases}. We expect the Shklovskii and Galactic contributions to have similar orders of magnitude but of opposite sign, likely cancelling each other out. However, the exact values will only be computable when precise measurements of PM and $d$ are available, and it may well be possible that either of them is dominant over the other. Comparing it with the values listed in Table~\ref{measurements} ($\dot P=2.6\pm1.0$~s\,s$^{-1}$), we see that the measured $\dot P$ may also be heavily contaminated by either contribution. Nonetheless, it is also clear that our mass constraints should not be biased by the proper motion contribution to $\dot\omega$, as the maximum expected contribution is still three orders of magnitude smaller than our current measurement uncertainty, while the effect on $\dot P_\textrm{b}$ will eventually be able to constrain the distance through comparison with the prediction given by GR (Table~\ref{measurements}, $\dot P_\textrm{b}^{\textrm{GR}}=1.225^{+0.026}_{-0.009}$~s\,s$^{-1}$).

\subsection{Future prospects for timing}\label{prospects_estimation}

\begin{figure}
\centering
 \includegraphics[width=\columnwidth]{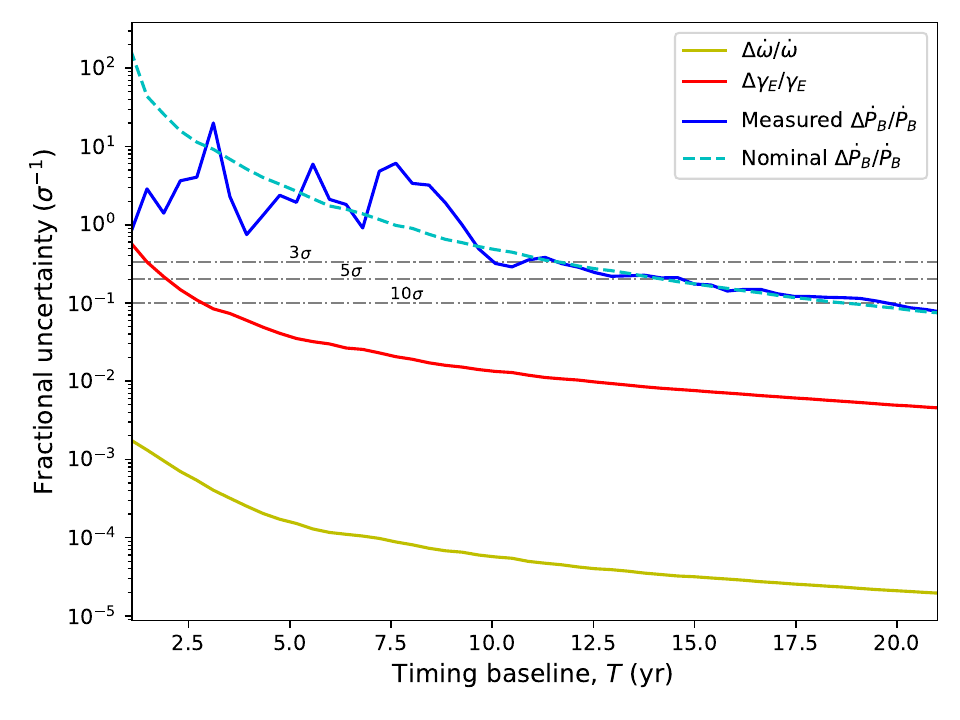}
 \caption{Fractional uncertainty evolution with timing baseline of the measured $\dot\omega$, $\dot\gamma_\textrm{E}$ and $\dot P_\textrm{b}$ PK parameters from the \texttt{TEMPO2/fake} simulation (solid lines: uncertainty/measurement, cyan dashed line: uncertainty/simulation input value, grey dashed lines: significance thresholds). For $\dot P_\textrm{b}$, the dashed line is taken as a reference due to the low significance of the measurement. $\dot\omega$ and $\dot\gamma_\textrm{E}$ scale with $T^{-3/2}$, while $\dot P_\textrm{b}$ scales with $T^{-5/2}$.}
 \label{significance}
\end{figure}

We estimate the prospects for improved mass measurements of J1208$-$5936 by simulating new ToAs with the \texttt{TEMPO2/fake} plug-in. We emulate the observing cadence outside of the Shapiro delay campaign (4 ToAs/month), our current timing precision (37~\textmu s) and add an assumed orbital decay of $\dot P_\textrm{b}=-1.247533\times10^{-13}$~s/s from the crossing of the current values of $\dot\omega$ and $\gamma_\textrm{E}$. We also keep $h_3$ and $\varsigma$ fixed to their current values. As seen in Fig.~\ref{mass_diagram}, $\dot\omega$ is already providing a very tight constraint, so the evolution of the uncertainties on $M_\textrm{p}$, $M_\textrm{c}$ and $i$ is expected to be dependant on the evolution of the $\gamma_\textrm{E}$ measurement. From Fig.~\ref{significance}, we expect the precision of $\gamma_\textrm{E}$ to surpass a significance of 10\textsigma\, at $T=3$~yr, decreasing the individual uncertainties of $M_\textrm{p}$ and $M_\textrm{c}$ to $\pm0.1$~M$_\odot$, and that of $i$ to $\pm5$~deg. Beyond this point, the uncertainties on PK parameters are likely to become dominated by correlated spin and DM noise, but assuming only white noise or a good modelling with Bayesian sampling of correlated noise as offered by \texttt{TempoNEST} \citep{lentati2014temponest}, the first reliable measurement of $\dot P_\textrm{b}$ at 3\textsigma\, is reached at $T=11$~yr, while 5\textsigma\, is reached at $T=14$~yr, and 10\textsigma\, at $T=19$~yr. At this time, the effects of Shklovskii and the Galactic acceleration field will already have an impact on the measurement, which would only be accountable from a good measurement of the PM. Depending on the magnitude of the PM, an independent estimate of the distance may be performed at that time. The uncertainties on the masses at those times are $\pm0.01$, $\pm0.008$ and $\pm0.005$~M$_\odot$, and for $i$ they are $\pm1$~deg, $\pm0.5$~deg and $\pm0.4$~deg. The parameters $\dot\omega$ and $\gamma_\textrm{E}$ always remain the most constraining ones, with $\gamma_\textrm{E}$ dominating the uncertainty and $\dot P_\textrm{b}$ providing an independent estimate of true distance by forcing it to be consistent with GR and the measured PM, which in turn will help constrain the true value of $\dot P$.

It is unlikely that the geometry of the system will be well determined in the future. Even at $T=20$~yr, the proper motion contribution to $\dot\omega$ (Equation~\ref{omdotPM}, Table~\ref{biases}) is expected to have a maximum possible value which is one order of magnitude below the uncertainty with our assumed PM magnitude. $\dot x$ is also unlikely to be constrained in the following decades, as it is highly degenerate with $\gamma_\textrm{E}$, resulting in a spurious measurement until the precession angle of $\omega$ is large enough to break the degeneracy \citep[for a discussion of this phenomenon, the reader is encouraged to consult][]{ridolfi2019possible}. Furthermore, the PM contribution to $\dot x$ (equation~\ref{xdotPM}, Table~\ref{biases}) is expected to have a maximum possible value one hundred times smaller than the GR-predicted one (Table~\ref{measurements}). Therefore, a measurement of the ascending node $\Omega_\textrm{a}$ and a breaking of the sign degeneracy of $i$ will remain unlikely unless a large, unexpected PM magnitude is detected.

\subsection{Future profile changes}

The two components present in the profile (Section~\ref{profile_section}, Fig.~\ref{components}) are likely to change both in relative amplitude and in phase separation due to the geodetic precession of the pulsar spin axis around the orbital angular momentum vector during the following years \citep{damour1974precession}. In GR and assuming $M_\textrm{p}=M_\textrm{c}$, the rate of the spin-orbit coupling-induced precession is proportional to $\dot\omega$ \citep{barker1975gravitational} as in
\begin{equation}\label{geodetic}\Omega_\textrm{g}=\frac{7}{24}\dot\omega\textrm{,}\end{equation}
which in the case of J1208$-$5936 results in an expected value of $\Omega_g\approx0.268$~deg\,yr$^{-1}$. Assuming a cone-shaped pulse from which we are seeing the maximal cross-section, a spin axis perpendicular to the angular momentum and the pulse being emitted from the equator, this would give us a minimum timescale of $\approx$24~yr before the pulses precess out of view. However, a different spin-orbit orientation would only extend this range upwards and the primary component will most likely be visible for a longer period of time. A detailed study of profile changes during the following years will enable a more accurate prediction from the detection of any change in the profile or lack thereof.

\section{Conclusions}

In this work we report the discovery and follow-up study of the MMGPS-L discovery J1208$-$5936. Spinning at 28.71 ms and in close orbit with another NS, this is in the tenth known Galactic DNS merging within the Hubble time. We have constrained the masses and inclination angle to $M_\textrm{p}=1.26^{+0.13}_{-0.25}$~M$_\odot$, $M_\textrm{c}=1.32^{+0.25}_{-0.13}$~M$_\odot$ and $i=57\pm12$~deg from the mapping $\chi^2$ mapping of DDGR solutions, with the tightest constraint coming from the 900\textsigma\, measurement of the periastron advance $\dot\omega$. The measurement of $\dot P<4\times10^{-4}$~s\,s$^{-1}$ is consistent with a mildly recycled pulsar and makes J1208$-$5936 the pulsar in a DNS with the smallest period derivative. However, the value is likely to be biased by the Shklovskii and Galactic acceleration field contamination. Its high eccentricity is still consistent with the tail of eccentricity distribution arising from a 50~km\,s$^{-1}$ supernova kick during the formation of the companion NS \citep{tauris2017formation}, but at the same time it could be indicative of a larger supernova kick caused by a massive He star. A more precise measurement of $M_\textrm{c}$ in the future may clarify which is the case and help confirm the idea of two main formation channels for Galactic DNS systems depending on the supernova type. We have been unable to detect polarised emission from J1208$-$5936, but we have observed a faint, secondary leading component to the main pulse that becomes more prominent at high frequencies, and significantly scattered at low frequencies. We have also found robust evidence for scattering, with a scattering index of $2.8\pm0.2$, even though a better modelling may be able to provide a better measurement of the scattering index.

The merger time of $\tau_\textrm{m}=7.2\pm0.2$~Gyr adds J1208$-$5936 to the family of DNS systems merging withing the Hubble time, therefore making it a progenitor of NS merger events seen by gravitational-wave observatories such as the landmark GW170817 event \citep{abbott2017gw170817}, making it relevant for predictions of the cosmic NS merger rate based on Galactic binaries. The performance of the MMGPS-L, the most sensitive survey in the southern sky, encouraged us to revisit these predictions. The end result provides an updated merger rate of $\pazocal{R}_\textrm{MW}^\textrm{new}=25^{+19}_{-9}$~Myr$^{-1}$ and local cosmic merger rate of $\pazocal{R}_\textrm{local}^\textrm{new}=293^{+222}_{-103}$~Gpc$^{-3}$\,yr$^{-1}$ within a 90\% confidence interval, smaller than the limits provided by previous studies on Galactic DNS systems \citep{grunthal2021revisiting,pol2019future,pol2020updated} owing to the fact that the despite the high sensitivity of the MMGPS-L only one new system merging within the Hubble time has been discovered, reducing the expected number of unseen DNS systems. This continues the trend of more constrained, decreasing estimates over time as the depth of pulsar surveys increase and the modelling of pulsar beam shapes improves. The resulting prediction for the LIGO-Virgo-KAGRA O4 run is the observation of $10^{+8}_{-4}$ NS merger events within 90\% credible intervals.

We expect the mass constraints in this system to improve significantly in the following years. Through simulations, we predict the masses and inclination angle uncertainty to be reduced to $\pm0.1$~M$_\odot$ and $\pm5$~deg with only two extra years of timing. After two decades, mass and inclination angle uncertainties can be reduced down to $\pm0.005$~M$_\odot$ and $\pm0.4$~deg, with the uncertainty always being dominated by the precision in the measurement of the Einstein delay amplitude. An eventual detection of $\dot P_\textrm{b}$ will help constrain the true distance to the system by forcing consistency with GR and the PM.

Deep surveys with new sensitive facilities such as MeerKAT, FAST or the SKA in the future will continue to provide new systems similar to J1208$-$5936, and increase the discovery rate of DNS systems with improved sensitivity and search algorithms. The MPIfR-MeerKAT Galactic Plane survey at S-band \citep{kramer2016meerkat,padmanabh2023mmgps-l} will probe deep into southern Galactic plane at high radio frequencies, allowing the discovery of even more distant and faint compact pulsar binaries without being hampered by propagation effects introduced by the interstellar medium. Further in time, space-based gravitational-wave observatories like LISA will to probe tens of electromagnetically dark DNS systems with orbital periods of one hour or less \citep{lau2020detecting}, constraining estimates of the merger rate in the Milky Way even further.

\begin{acknowledgements}
The MeerKAT telescope is operated by the South African Radio Astronomy Observatory, which is a facility of the National Research Foundation, an agency of the Department of Science and Innovation. The Parkes radio telescope is part of the ATNF, which is funded by the Australian Government for operation as a National Facility managed by the Commonwealth Scientific and Industrial Research Organisation. We acknowledge the Wiradjuri people as the Traditional Owners of the Observatory site. SARAO acknowledges the ongoing advice and calibration of GPS systems by the National Metrology Institute of South Africa (NMISA) and the time space reference systems department of the Paris Observatory. Observations used the FBFUSE and APSUSE computing clusters for data acquisition, storage and analysis. These clusters were funded and installed by the MPIfR and the Max-Planck-Gesellschaft (MPG). All authors affiliated with the MPG acknowledge its constant support. Marina Berezina acknowledges support from the Bundesministerium für Bildung und Forschung D-MeerKAT award 05A17VH3 (Verbundprojekt D-MeerKAT). Marta Burgay acknowledges support through the research grant 'iPeska' (PI: A. Possenti) funded under the INAF national call Prin-SKA/CTA approved with the Presidential Decree 70/2016. Vivek Venkatraman Krishnan acknowledges financial support from the European Research Council (ERC) starting grant 'COMPACT' (grant agreement number: 101078094). We also thank Alessandro Ridolfi for providing a working version of \texttt{pysolator.py} and for his input in Section~\ref{companion_search}, Norbert Wex for his comments on the interpretations of the NS merger rate results in Section~\ref{merger_implications}, and Livia Silva Rocha and Robert Main for their general feedback on this manuscript. The data underlying this work will be shared on reasonable request to the MMGPS collaboration.
\end{acknowledgements}

%%%%%%%%%%%%%%%%%%%%%%%%%%%%%%%%%%%%%%%%%%%%%%%%%%
%\section*{Data Availability}
 
%The data underlying this article will be shared on reasonable request to the MMGPS collaboration.


\bibliography{J1208.bib}

\begin{thebibliography}{97}
\expandafter\ifx\csname natexlab\endcsname\relax\def\natexlab#1{#1}\fi

\bibitem[{{Abbott} {et~al.}(2017{\natexlab{a}}){Abbott}, {Abbott}, {Abbott},
  {Acernese}, {Ackley}, {Adams}, {Adams}, {Addesso}, {Adhikari}, {Adya},
  {Affeldt}, {Afrough}, {Agarwal}, {Agathos}, {Agatsuma}, {Aggarwal}, {Aguiar},
  {Aiello}, {Ain}, {Ajith}, {Allen}, {Allen}, {Allocca}, {Altin}, {Amato},
  {Ananyeva}, {Anderson}, {Anderson}, {Angelova}, {Antier}, {Appert}, {Arai},
  {Araya}, {Areeda}, {Arnaud}, {Arun}, {Ascenzi}, {Ashton}, {Ast}, {Aston},
  {Astone}, {Atallah}, {Aufmuth}, {et~al.}}]{abbott2017multi}
{Abbott}, B.~P., {Abbott}, R., {Abbott}, T.~D., {et~al.} 2017{\natexlab{a}},
  \apjl, 848, L12

\bibitem[{{Abbott} {et~al.}(2017{\natexlab{b}}){Abbott}, {Abbott}, {Abbott},
  {Acernese}, {Ackley}, {Adams}, {Adams}, {Addesso}, {Adhikari}, {Adya},
  {Affeldt}, {Afrough}, {Agarwal}, {Agathos}, {Agatsuma}, {Aggarwal}, {Aguiar},
  {Aiello}, {Ain}, {Ajith}, {Allen}, {Allen}, {Allocca}, {Altin}, {Amato},
  {Ananyeva}, {Anderson}, {Anderson}, {Angelova},
  {et~al.}}]{abbott2017gw170817}
{Abbott}, B.~P., {Abbott}, R., {Abbott}, T.~D., {et~al.} 2017{\natexlab{b}},
  \prl, 119, 161101

\bibitem[{{Abbott} {et~al.}(2021){Abbott}, {Abbott}, {Acernese}, {Ackley},
  {Adams}, {Adhikari}, {Adhikari}, {Adya}, {Affeldt}, {Agarwal}, {Agathos},
  {Agatsuma}, {Aggarwal}, {Aguiar}, {Aiello}, {Ain}, {Ajith}, {Albanesi},
  {Allocca}, {Altin}, {Amato}, {Anand}, {Anand}, {Ananyeva}, {Anderson},
  {Anderson}, {Andrade}, {Andres}, {Andri{\'c}}, {et~al.}}]{abbott2021gwtc}
{Abbott}, R., {Abbott}, T.~D., {Acernese}, F., {et~al.} 2021, arXiv e-prints,
  arXiv:2108.01045

\bibitem[{{Agazie} {et~al.}(2021){Agazie}, {Mingyar}, {McLaughlin}, {Swiggum},
  {Kaplan}, {Blumer}, {Chawla}, {DeCesar}, {Demorest}, {Fiore}, {Fonseca},
  {Gelfand}, {Kaspi}, {Kondratiev}, {LaRose}, {van Leeuwen}, {Levin}, {Lewis},
  {Lynch}, {McEwen}, {Al Noori}, {Parent}, {Ransom}, {Roberts}, {Schmiedekamp},
  {Schmiedekamp}, {Siemens}, {Spiewak}, {Stairs}, \&
  {Surnis}}]{agazie2021green}
{Agazie}, G.~Y., {Mingyar}, M.~G., {McLaughlin}, M.~A., {et~al.} 2021, \apj,
  922, 35

\bibitem[{{Allen} {et~al.}(2013){Allen}, {Knispel}, {Cordes}, {Deneva},
  {Hessels}, {Anderson}, {Aulbert}, {Bock}, {Brazier}, {Chatterjee},
  {Demorest}, {Eggenstein}, {Fehrmann}, {Gotthelf}, {Hammer}, {Kaspi},
  {Kramer}, {Lyne}, {Machenschalk}, {McLaughlin}, {Messenger}, {Pletsch},
  {Ransom}, {Stairs}, {Stappers}, {Bhat}, {Bogdanov}, {Camilo}, {Champion},
  {Crawford}, {Desvignes}, {Freire}, {Heald}, {Jenet}, {Lazarus}, {Lee}, {van
  Leeuwen}, {Lynch}, {Papa}, {Prix}, {Rosen}, {Scholz}, {Siemens}, {Stovall},
  {Venkataraman}, \& {Zhu}}]{allen2013einstein}
{Allen}, B., {Knispel}, B., {Cordes}, J.~M., {et~al.} 2013, \apj, 773, 91

\bibitem[{{Andersen} \& {Ransom}(2018)}]{andersen2018fourier}
{Andersen}, B.~C. \& {Ransom}, S.~M. 2018, \apjl, 863, L13

\bibitem[{{Andrews} \& {Mandel}(2019)}]{andrews2019double}
{Andrews}, J.~J. \& {Mandel}, I. 2019, \apjl, 880, L8

\bibitem[{{Bagchi} {et~al.}(2013){Bagchi}, {Lorimer}, \&
  {Wolfe}}]{bagchi2013detectability}
{Bagchi}, M., {Lorimer}, D.~R., \& {Wolfe}, S. 2013, \mnras, 432, 1303

\bibitem[{{Bailes} {et~al.}(2020){Bailes}, {Jameson}, {Abbate}, {Barr}, {Bhat},
  {Bondonneau}, {Burgay}, {Buchner}, {Camilo}, {Champion}, {Cognard},
  {Demorest}, {Freire}, {Gautam}, {Geyer}, {Griessmeier}, {Guillemot}, {Hu},
  {Jankowski}, {Johnston}, {Karastergiou}, {Karuppusamy}, {Kaur}, {Keith},
  {Kramer}, {van Leeuwen}, {Lower}, {Maan}, {McLaughlin}, {Meyers},
  {Os{\l}owski}, {Oswald}, {Parthasarathy}, {Pennucci}, {Posselt}, {Possenti},
  {Ransom}, {Reardon}, {Ridolfi}, {Schollar}, {Serylak}, {Shaifullah},
  {Shamohammadi}, {Shannon}, {Sobey}, {Song}, {Spiewak}, {Stairs}, {Stappers},
  {van Straten}, {Szary}, {Theureau}, {Venkatraman Krishnan}, {Weltevrede},
  {Wex}, {Abbott}, {Adams}, {Burger}, {Gamatham}, {Gouws}, {Horn}, {Hugo},
  {Joubert}, {Manley}, {McAlpine}, {Passmoor}, {Peens-Hough}, {Ramudzuli},
  {Rust}, {Salie}, {Schwardt}, {Siebrits}, {Van Tonder}, {Van Tonder}, \&
  {Welz}}]{bailes2020meerkat}
{Bailes}, M., {Jameson}, A., {Abbate}, F., {et~al.} 2020, \pasa, 37, e028

\bibitem[{{Balakrishnan} {et~al.}(2022){Balakrishnan}, {Champion}, {Barr},
  {Kramer}, {Venkatraman Krishnan}, {Eatough}, {Sengar}, \&
  {Bailes}}]{balakrishnan2022coherent}
{Balakrishnan}, V., {Champion}, D., {Barr}, E., {et~al.} 2022, \mnras, 511,
  1265

\bibitem[{{Barker} \& {O'Connell}(1975)}]{barker1975gravitational}
{Barker}, B.~M. \& {O'Connell}, R.~F. 1975, \prd, 12, 329

\bibitem[{{Barr}(2020)}]{barr2020peasoup}
{Barr}, E. 2020, {Peasoup: C++/CUDA GPU pulsar searching library}, Astrophysics
  Source Code Library, record ascl:2001.014

\bibitem[{{Bhattacharya} \& {van den Heuvel}(1991)}]{bhattacharya1991formation}
{Bhattacharya}, D. \& {van den Heuvel}, E.~P.~J. 1991, \physrep, 203, 1

\bibitem[{{Bhattacharyya} \&
  {Nityananda}(2008)}]{bhattacharyya2008determination}
{Bhattacharyya}, B. \& {Nityananda}, R. 2008, \mnras, 387, 273

\bibitem[{{Burgay} {et~al.}(2003){Burgay}, {D'Amico}, {Possenti}, {Manchester},
  {Lyne}, {Joshi}, {McLaughlin}, {Kramer}, {Sarkissian}, {Camilo}, {Kalogera},
  {Kim}, \& {Lorimer}}]{burgay2006j0737}
{Burgay}, M., {D'Amico}, N., {Possenti}, A., {et~al.} 2003, \nat, 426, 531

\bibitem[{{Burgay} {et~al.}(2006){Burgay}, {Joshi}, {D'Amico}, {Possenti},
  {Lyne}, {Manchester}, {McLaughlin}, {Kramer}, {Camilo}, \&
  {Freire}}]{burgay2006parkes}
{Burgay}, M., {Joshi}, B.~C., {D'Amico}, N., {et~al.} 2006, \mnras, 368, 283

\bibitem[{{Cameron} {et~al.}(2023){Cameron}, {Bailes}, {Champion}, {Freire},
  {Kramer}, {McLaughlin}, {Ng}, {Possenti}, {Ridolfi}, {Tauris}, {Wahl}, \&
  {Wex}}]{cameron2023relativistic}
{Cameron}, A.~D., {Bailes}, M., {Champion}, D.~J., {et~al.} 2023, \mnras, 523,
  5064

\bibitem[{{Cameron} {et~al.}(2017){Cameron}, {Barr}, {Champion}, {Kramer}, \&
  {Zhu}}]{cameron2017investigation}
{Cameron}, A.~D., {Barr}, E.~D., {Champion}, D.~J., {Kramer}, M., \& {Zhu},
  W.~W. 2017, \mnras, 468, 1994

\bibitem[{{Chattopadhyay} {et~al.}(2020){Chattopadhyay}, {Stevenson}, {Hurley},
  {Rossi}, \& {Flynn}}]{chattopadhyay2020modelling}
{Chattopadhyay}, D., {Stevenson}, S., {Hurley}, J.~R., {Rossi}, L.~J., \&
  {Flynn}, C. 2020, \mnras, 494, 1587

\bibitem[{Chen {et~al.}(2021)Chen, Holz, Miller, Evans, Vitale, \&
  Creighton}]{chen2021distance}
Chen, H.-Y., Holz, D.~E., Miller, J., {et~al.} 2021, Class. Quantum Gravity,
  38, 055010

\bibitem[{{Chen} {et~al.}(2021){Chen}, {Barr}, {Karuppusamy}, {Kramer}, \&
  {Stappers}}]{chen2021wide}
{Chen}, W., {Barr}, E., {Karuppusamy}, R., {Kramer}, M., \& {Stappers}, B.
  2021, J. Astron. Instrum., 10, 2150013

\bibitem[{{Cordes}(2004)}]{cordes2002ne2001}
{Cordes}, J.~M. 2004, ASP Conference Series, 317, 211

\bibitem[{{Cordes} {et~al.}(2006){Cordes}, {Freire}, {Lorimer}, {Camilo},
  {Champion}, {Nice}, {Ramachandran}, {Hessels}, {Vlemmings}, {van Leeuwen},
  {Ransom}, {Bhat}, {Arzoumanian}, {McLaughlin}, {Kaspi}, {Kasian}, {Deneva},
  {Reid}, {Chatterjee}, {Han}, {Backer}, {Stairs}, {Deshpande}, \&
  {Faucher-Gigu{\`e}re}}]{cordes2006arecibo}
{Cordes}, J.~M., {Freire}, P.~C.~C., {Lorimer}, D.~R., {et~al.} 2006, \apj,
  637, 446

\bibitem[{{Corongiu} {et~al.}(2007){Corongiu}, {Kramer}, {Stappers}, {Lyne},
  {Jessner}, {Possenti}, {D'Amico}, \& {L{\"o}hmer}}]{corongiu2007binary}
{Corongiu}, A., {Kramer}, M., {Stappers}, B.~W., {et~al.} 2007, \aap, 462, 703

\bibitem[{Damour \& Ruffini(1974)}]{damour1974precession}
Damour, .~T. \& Ruffini, R. 1974, C. R. Acad. Sc. Paris, Serie A, 279, 971

\bibitem[{{Damour} \& {Deruelle}(1986)}]{damour1986general}
{Damour}, T. \& {Deruelle}, N. 1986, Ann. Inst. Henri Poincar{\'e} Phys.
  Th{\'e}or, 44, 263

\bibitem[{{Damour} \& {Taylor}(1991)}]{damour2992orbital}
{Damour}, T. \& {Taylor}, J.~H. 1991, \apj, 366, 501

\bibitem[{{Dewi} {et~al.}(2005){Dewi}, {Podsiadlowski}, \&
  {Pols}}]{dewi2005spin}
{Dewi}, J.~D.~M., {Podsiadlowski}, P., \& {Pols}, O.~R. 2005, \mnras, 363, L71

\bibitem[{{Eatough} {et~al.}(2021){Eatough}, {Torne}, {Desvignes}, {Kramer},
  {Karuppusamy}, {Klein}, {Spitler}, {Lee}, {Champion}, {Liu}, {Wharton},
  {Rezzolla}, \& {Falcke}}]{eatough2021effelsberg}
{Eatough}, R.~P., {Torne}, P., {Desvignes}, G., {et~al.} 2021, \mnras, 507,
  5053

\bibitem[{{Edwards} {et~al.}(2006){Edwards}, {Hobbs}, \&
  {Manchester}}]{edwards2006tempo2}
{Edwards}, R.~T., {Hobbs}, G.~B., \& {Manchester}, R.~N. 2006, \mnras, 372,
  1549

\bibitem[{{Faulkner} {et~al.}(2005{\natexlab{a}}){Faulkner}, {Kramer}, {Lyne},
  {Manchester}, {McLaughlin}, {Stairs}, {Hobbs}, {Possenti}, {Lorimer},
  {D'Amico}, {Camilo}, \& {Burgay}}]{faulkner2005new}
{Faulkner}, A.~J., {Kramer}, M., {Lyne}, A.~G., {et~al.} 2005{\natexlab{a}},
  \apjl, 618, L119

\bibitem[{{Faulkner} {et~al.}(2005{\natexlab{b}}){Faulkner}, {Kramer}, {Lyne},
  {Manchester}, {McLaughlin}, {Stairs}, {Hobbs}, {Possenti}, {Lorimer},
  {D'Amico}, {Camilo}, \& {Burgay}}]{faulkner2004psr}
{Faulkner}, A.~J., {Kramer}, M., {Lyne}, A.~G., {et~al.} 2005{\natexlab{b}},
  \apjl, 618, L119

\bibitem[{{Ferdman} {et~al.}(2020){Ferdman}, {Freire}, {Perera}, {Pol},
  {Camilo}, {Chatterjee}, {Cordes}, {Crawford}, {Hessels}, {Kaspi},
  {McLaughlin}, {Parent}, {Stairs}, \& {van Leeuwen}}]{ferdman2020asymmetric}
{Ferdman}, R.~D., {Freire}, P.~C.~C., {Perera}, B.~B.~P., {et~al.} 2020, \nat,
  583, 211

\bibitem[{{Ferdman} {et~al.}(2014){Ferdman}, {Stairs}, {Kramer}, {Janssen},
  {Bassa}, {Stappers}, {Demorest}, {Cognard}, {Desvignes}, {Theureau},
  {Burgay}, {Lyne}, {Manchester}, \& {Possenti}}]{ferdman2014psr}
{Ferdman}, R.~D., {Stairs}, I.~H., {Kramer}, M., {et~al.} 2014, \mnras, 443,
  2183

\bibitem[{{Freire} \& {Ridolfi}(2018)}]{freire2018algorithm}
{Freire}, P. C.~C. \& {Ridolfi}, A. 2018, \mnras, 476, 4794

\bibitem[{{Freire} \& {Wex}(2010)}]{freire2010orthometric}
{Freire}, P. C.~C. \& {Wex}, N. 2010, \mnras, 409, 199

\bibitem[{{Grunthal} {et~al.}(2021){Grunthal}, {Kramer}, \&
  {Desvignes}}]{grunthal2021revisiting}
{Grunthal}, K., {Kramer}, M., \& {Desvignes}, G. 2021, \mnras, 507, 5658

\bibitem[{{Haniewicz} {et~al.}(2021){Haniewicz}, {Ferdman}, {Freire},
  {Champion}, {Bunting}, {Lorimer}, \& {McLaughlin}}]{haniewicz2021precise}
{Haniewicz}, H.~T., {Ferdman}, R.~D., {Freire}, P.~C.~C., {et~al.} 2021,
  \mnras, 500, 4620

\bibitem[{{Hobbs} {et~al.}(2020){Hobbs}, {Manchester}, {Dunning}, {Jameson},
  {Roberts}, {George}, {Green}, {Tuthill}, {Toomey}, {Kaczmarek}, {Mader},
  {Marquarding}, {Ahmed}, {Amy}, {Bailes}, {Beresford}, {Bhat}, {Bock},
  {Bourne}, {Bowen}, {Brothers}, {Cameron}, {Carretti}, {Carter}, {Castillo},
  {Chekkala}, {Cheng}, {Chung}, {Craig}, {Dai}, {Dawson}, {Dempsey}, {Doherty},
  {Dong}, {Edwards}, {Ergesh}, {Gao}, {Han}, {Hayman}, {Indermuehle},
  {Jeganathan}, {Johnston}, {Kanoniuk}, {Kesteven}, {Kramer}, {Leach},
  {Mcintyre}, {Moss}, {Os{\l}owski}, {Phillips}, {Pope}, {Preisig}, {Price},
  {Reeves}, {Reilly}, {Reynolds}, {Robishaw}, {Roush}, {Ruckley}, {Sadler},
  {Sarkissian}, {Severs}, {Shannon}, {Smart}, {Smith}, {Smith}, {Sobey},
  {Staveley-Smith}, {Tzioumis}, {van Straten}, {Wang}, {Wen}, \&
  {Whiting}}]{george2020uwl}
{Hobbs}, G., {Manchester}, R.~N., {Dunning}, A., {et~al.} 2020, \pasa, 37, e012

\bibitem[{{Hobbs} {et~al.}(2006){Hobbs}, {Edwards}, \&
  {Manchester}}]{hobbs2006tempo2}
{Hobbs}, G.~B., {Edwards}, R.~T., \& {Manchester}, R.~N. 2006, \mnras, 369, 655

\bibitem[{{Hotan} {et~al.}(2004){Hotan}, {van Straten}, \&
  {Manchester}}]{hotan2004psrchive}
{Hotan}, A.~W., {van Straten}, W., \& {Manchester}, R.~N. 2004, \pasa, 21, 302

\bibitem[{{Hu} {et~al.}(2020){Hu}, {Kramer}, {Wex}, {Champion}, \&
  {Kehl}}]{hu2020constraining}
{Hu}, H., {Kramer}, M., {Wex}, N., {Champion}, D.~J., \& {Kehl}, M.~S. 2020,
  \mnras, 497, 3118

\bibitem[{{Janssen} {et~al.}(2008){Janssen}, {Stappers}, {Kramer}, {Nice},
  {Jessner}, {Cognard}, \& {Purver}}]{janssen2008multi}
{Janssen}, G.~H., {Stappers}, B.~W., {Kramer}, M., {et~al.} 2008, \aap, 490,
  753

\bibitem[{{Johnston} \& {Kulkarni}(1991)}]{johnston1991detectability}
{Johnston}, H.~M. \& {Kulkarni}, S.~R. 1991, \apj, 368, 504

\bibitem[{{Jonas} \& {MeerKAT Team}(2016)}]{jonas2016meerkat}
{Jonas}, J. \& {MeerKAT Team}. 2016, in MeerKAT Science: On the Pathway to the
  SKA, 1

\bibitem[{{Kalogera} {et~al.}(2004){Kalogera}, {Kim}, {Lorimer}, {Burgay},
  {D'Amico}, {Possenti}, {Manchester}, {Lyne}, {Joshi}, {McLaughlin}, {Kramer},
  {Sarkissian}, \& {Camilo}}]{kalogera2004cosmic}
{Kalogera}, V., {Kim}, C., {Lorimer}, D.~R., {et~al.} 2004, \apjl, 601, L179

\bibitem[{{Keith} {et~al.}(2010){Keith}, {Jameson}, {van Straten}, {Bailes},
  {Johnston}, {Kramer}, {Possenti}, {Bates}, {Bhat}, {Burgay}, {Burke-Spolaor},
  {D'Amico}, {Levin}, {McMahon}, {Milia}, \& {Stappers}}]{keith2010high}
{Keith}, M.~J., {Jameson}, A., {van Straten}, W., {et~al.} 2010, \mnras, 409,
  619

\bibitem[{{Keith} {et~al.}(2009){Keith}, {Kramer}, {Lyne}, {Eatough}, {Stairs},
  {Possenti}, {Camilo}, \& {Manchester}}]{keith2009psr}
{Keith}, M.~J., {Kramer}, M., {Lyne}, A.~G., {et~al.} 2009, \mnras, 393, 623

\bibitem[{{Kim} {et~al.}(2010){Kim}, {Kalogera}, \& {Lorimer}}]{kim2010effect}
{Kim}, C., {Kalogera}, V., \& {Lorimer}, D. 2010, \nar, 54, 148

\bibitem[{{Kim} {et~al.}(2003){Kim}, {Kalogera}, \&
  {Lorimer}}]{kim2003probability}
{Kim}, C., {Kalogera}, V., \& {Lorimer}, D.~R. 2003, \apj, 584, 985

\bibitem[{{Kim} {et~al.}(2015){Kim}, {Perera}, \&
  {McLaughlin}}]{kim2015implications}
{Kim}, C., {Perera}, B. B.~P., \& {McLaughlin}, M.~A. 2015, \mnras, 448, 928

\bibitem[{{Kopeikin}(1996)}]{kopeikin1996proper}
{Kopeikin}, S.~M. 1996, \apjl, 467, L93

\bibitem[{{Kopparapu} {et~al.}(2008){Kopparapu}, {Hanna}, {Kalogera},
  {O'Shaughnessy}, {Gonz{\'a}lez}, {Brady}, \& {Fairhurst}}]{kopparapu2008host}
{Kopparapu}, R.~K., {Hanna}, C., {Kalogera}, V., {et~al.} 2008, \apj, 675, 1459

\bibitem[{{Kramer} {et~al.}(2016){Kramer}, {Menten}, {Barr}, {Karuppusamy},
  {Kasemann}, {Klein}, {Ros}, {Wieching}, \& {Wucknitz}}]{kramer2016meerkat}
{Kramer}, M., {Menten}, K., {Barr}, E.~D., {et~al.} 2016, in MeerKAT Science:
  On the Pathway to the SKA, 3

\bibitem[{{Kramer} {et~al.}(2021){Kramer}, {Stairs}, {Manchester}, {Wex},
  {Deller}, {Coles}, {Ali}, {Burgay}, {Camilo}, {Cognard}, {Damour},
  {Desvignes}, {Ferdman}, {Freire}, {Grondin}, {Guillemot}, {Hobbs}, {Janssen},
  {Karuppusamy}, {Lorimer}, {Lyne}, {McKee}, {McLaughlin}, {M{\"u}nch},
  {Perera}, {Pol}, {Possenti}, {Sarkissian}, {Stappers}, \&
  {Theureau}}]{kramer2021strong}
{Kramer}, M., {Stairs}, I.~H., {Manchester}, R.~N., {et~al.} 2021, Physical
  Review X, 11, 041050

\bibitem[{{Lau} {et~al.}(2020){Lau}, {Mandel}, {Vigna-G{\'o}mez}, {Neijssel},
  {Stevenson}, \& {Sesana}}]{lau2020detecting}
{Lau}, M. Y.~M., {Mandel}, I., {Vigna-G{\'o}mez}, A., {et~al.} 2020, \mnras,
  492, 3061

\bibitem[{{Lentati} {et~al.}(2014){Lentati}, {Alexander}, {Hobson}, {Feroz},
  {van Haasteren}, {Lee}, \& {Shannon}}]{lentati2014temponest}
{Lentati}, L., {Alexander}, P., {Hobson}, M.~P., {et~al.} 2014, \mnras, 437,
  3004

\bibitem[{Lorimer \& Kramer(2005)}]{lorimer2012handbook}
Lorimer, D.~R. \& Kramer, M. 2005, Handbook of pulsar astronomy (Cambridge
  University press)

\bibitem[{{Lynch} {et~al.}(2018){Lynch}, {Swiggum}, {Kondratiev}, {Kaplan},
  {Stovall}, {Fonseca}, {Roberts}, {Levin}, {DeCesar}, {Cui}, {Cenko},
  {Gatkine}, {Archibald}, {Banaszak}, {Biwer}, {Boyles}, {Chawla}, {Dartez},
  {Day}, {Ford}, {Flanigan}, {Hessels}, {Hinojosa}, {Jenet}, {Karako-Argaman},
  {Kaspi}, {Leake}, {Lunsford}, {Martinez}, {Mata}, {McLaughlin}, {Noori},
  {Ransom}, {Rohr}, {Siemens}, {Spiewak}, {Stairs}, {van Leeuwen}, {Walker}, \&
  {Wells}}]{lynch2018green}
{Lynch}, R.~S., {Swiggum}, J.~K., {Kondratiev}, V.~I., {et~al.} 2018, \apj,
  859, 93

\bibitem[{{Lyne} {et~al.}(2004){Lyne}, {Burgay}, {Kramer}, {Possenti},
  {Manchester}, {Camilo}, {McLaughlin}, {Lorimer}, {D'Amico}, {Joshi},
  {Reynolds}, \& {Freire}}]{lyne2004double}
{Lyne}, A.~G., {Burgay}, M., {Kramer}, M., {et~al.} 2004, Science, 303, 1153

\bibitem[{{Manchester} {et~al.}(2001){Manchester}, {Lyne}, {Camilo}, {Bell},
  {Kaspi}, {D'Amico}, {McKay}, {Crawford}, {Stairs}, {Possenti}, {Kramer}, \&
  {Sheppard}}]{manchester2001parkes}
{Manchester}, R.~N., {Lyne}, A.~G., {Camilo}, F., {et~al.} 2001, \mnras, 328,
  17

\bibitem[{{Martinez} {et~al.}(2015){Martinez}, {Stovall}, {Freire}, {Deneva},
  {Jenet}, {McLaughlin}, {Bagchi}, {Bates}, \& {Ridolfi}}]{martinez2015pulsar}
{Martinez}, J.~G., {Stovall}, K., {Freire}, P.~C.~C., {et~al.} 2015, \apj, 812,
  143

\bibitem[{{Martinez} {et~al.}(2017){Martinez}, {Stovall}, {Freire}, {Deneva},
  {Tauris}, {Ridolfi}, {Wex}, {Jenet}, {McLaughlin}, \&
  {Bagchi}}]{martinez2017pulsar}
{Martinez}, J.~G., {Stovall}, K., {Freire}, P.~C.~C., {et~al.} 2017, \apjl,
  851, L29

\bibitem[{{McMillan}(2017)}]{mcmillan2017mass}
{McMillan}, P.~J. 2017, \mnras, 465, 76

\bibitem[{{Morello} {et~al.}(2019){Morello}, {Barr}, {Cooper}, {Bailes},
  {Bates}, {Bhat}, {Burgay}, {Burke-Spolaor}, {Cameron}, {Champion}, {Eatough},
  {Flynn}, {Jameson}, {Johnston}, {Keith}, {Keane}, {Kramer}, {Levin}, {Ng},
  {Petroff}, {Possenti}, {Stappers}, {van Straten}, \&
  {Tiburzi}}]{morello2019high}
{Morello}, V., {Barr}, E.~D., {Cooper}, S., {et~al.} 2019, \mnras, 483, 3673

\bibitem[{{Ng} {et~al.}(2018){Ng}, {Kruckow}, {Tauris}, {Lyne}, {Freire},
  {Ridolfi}, {Caiazzo}, {Heyl}, {Kramer}, {Cameron}, {Champion}, \&
  {Stappers}}]{ng2018psr}
{Ng}, C., {Kruckow}, M.~U., {Tauris}, T.~M., {et~al.} 2018, \mnras, 476, 4315

\bibitem[{{Oswald} {et~al.}(2021){Oswald}, {Karastergiou}, {Posselt},
  {Johnston}, {Bailes}, {Buchner}, {Geyer}, {Keith}, {Kramer}, {Parthasarathy},
  {Reardon}, {Serylak}, {Shannon}, {Spiewak}, {van Straten}, \& {Venkatraman
  Krishnan}}]{oswald2021scatter}
{Oswald}, L.~S., {Karastergiou}, A., {Posselt}, B., {et~al.} 2021, \mnras, 504,
  1115

\bibitem[{{{\"O}zel} \& {Freire}(2016)}]{ozel2016masses}
{{\"O}zel}, F. \& {Freire}, P. 2016, \araa, 54, 401

\bibitem[{{Padmanabh} {et~al.}(2023){Padmanabh}, {Barr}, {Sridhar}, {Rugel},
  A., {et~al.}}]{padmanabh2023mmgps-l}
{Padmanabh}, P.~V., {Barr}, E.~D., {Sridhar}, S.~S., {et~al.} 2023, arXiv
  e-prints, arXiv:2303.09231

\bibitem[{{Peters}(1964)}]{peters1964gravitational}
{Peters}, P.~C. 1964, Physical Review, 136, 1224

\bibitem[{{Pol} {et~al.}(2019){Pol}, {McLaughlin}, \&
  {Lorimer}}]{pol2019future}
{Pol}, N., {McLaughlin}, M., \& {Lorimer}, D.~R. 2019, \apj, 870, 71

\bibitem[{{Pol} {et~al.}(2020){Pol}, {McLaughlin}, \&
  {Lorimer}}]{pol2020updated}
{Pol}, N., {McLaughlin}, M., \& {Lorimer}, D.~R. 2020, RNAAS, 4, 22

\bibitem[{{Ridolfi} {et~al.}(2019){Ridolfi}, {Freire}, {Gupta}, \&
  {Ransom}}]{ridolfi2019possible}
{Ridolfi}, A., {Freire}, P.~C.~C., {Gupta}, Y., \& {Ransom}, S.~M. 2019,
  \mnras, 490, 3860

\bibitem[{{Sengar} {et~al.}(2022){Sengar}, {Balakrishnan}, {Stevenson},
  {Bailes}, {Barr}, {Bhat}, {Burgay}, {Bernadich}, {Cameron}, {Champion},
  {Chen}, {Flynn}, {Jameson}, {Johnston}, {Keith}, {Kramer}, {Morello}, {Ng},
  {Possenti}, {Stappers}, {Shannon}, {van Straten}, \&
  {Wongphechauxsorn}}]{sengar2022high}
{Sengar}, R., {Balakrishnan}, V., {Stevenson}, S., {et~al.} 2022, \mnras, 512,
  5782

\bibitem[{{Shklovskii}(1970)}]{shklovskii1970possible}
{Shklovskii}, I.~S. 1970, \sovast, 13, 562

\bibitem[{{Singh} {et~al.}(2022){Singh}, {Roy}, {Panda}, {Bhattacharyya},
  {Morello}, {Stappers}, {Ray}, \& {McLaughlin}}]{singh2022gmrt}
{Singh}, S., {Roy}, J., {Panda}, U., {et~al.} 2022, \apj, 934, 138

\bibitem[{{Sofue}(2020)}]{sofue2020rotation}
{Sofue}, Y. 2020, Galaxies, 8, 37

\bibitem[{{Splaver} {et~al.}(2002){Splaver}, {Nice}, {Arzoumanian}, {Camilo},
  {Lyne}, \& {Stairs}}]{splaver2002masses}
{Splaver}, E.~M., {Nice}, D.~J., {Arzoumanian}, Z., {et~al.} 2002, \apj, 581,
  509

\bibitem[{{Staelin}(1969)}]{staelin1969fast}
{Staelin}, D.~H. 1969, IEEE Proceedings, 57, 724

\bibitem[{{Stappers} \& {Kramer}(2016)}]{stappers2016update}
{Stappers}, B. \& {Kramer}, M. 2016, in MeerKAT Science: On the Pathway to the
  SKA, 9

\bibitem[{{Stovall} {et~al.}(2018){Stovall}, {Freire}, {Chatterjee},
  {Demorest}, {Lorimer}, {McLaughlin}, {Pol}, {van Leeuwen}, {Wharton},
  {Allen}, {Boyce}, {Brazier}, {Caballero}, {Camilo}, {Camuccio}, {Cordes},
  {Crawford}, {Deneva}, {Ferdman}, {Hessels}, {Jenet}, {Kaspi}, {Knispel},
  {Lazarus}, {Lynch}, {Parent}, {Patel}, {Pleunis}, {Ransom}, {Scholz},
  {Seymour}, {Siemens}, {Stairs}, {Swiggum}, \& {Zhu}}]{stovall2018palfa}
{Stovall}, K., {Freire}, P.~C.~C., {Chatterjee}, S., {et~al.} 2018, \apjl, 854,
  L22

\bibitem[{{Stovall} {et~al.}(2014){Stovall}, {Lynch}, {Ransom}, {Archibald},
  {Banaszak}, {Biwer}, {Boyles}, {Dartez}, {Day}, {Ford}, {Flanigan}, {Garcia},
  {Hessels}, {Hinojosa}, {Jenet}, {Kaplan}, {Karako-Argaman}, {Kaspi},
  {Kondratiev}, {Leake}, {Lorimer}, {Lunsford}, {Martinez}, {Mata},
  {McLaughlin}, {Roberts}, {Rohr}, {Siemens}, {Stairs}, {van Leeuwen},
  {Walker}, \& {Wells}}]{stovall2014green}
{Stovall}, K., {Lynch}, R.~S., {Ransom}, S.~M., {et~al.} 2014, \apj, 791, 67

\bibitem[{{Suresh} {et~al.}(2022){Suresh}, {Cordes}, {Chatterjee}, {Gajjar},
  {Perez}, {Siemion}, {Lebofsky}, {MacMahon}, \& {Ng}}]{suresh2022gbt}
{Suresh}, A., {Cordes}, J.~M., {Chatterjee}, S., {et~al.} 2022, \apj, 933, 121

\bibitem[{{Swiggum} {et~al.}(2023){Swiggum}, {Pleunis}, {Parent}, {Kaplan},
  {McLaughlin}, {Stairs}, {Spiewak}, {Agazie}, {Chawla}, {DeCesar}, {Dolch},
  {Fiore}, {Fonseca}, {Istrate}, {Kaspi}, {Kondratiev}, {van Leeuwen}, {Levin},
  {Lewis}, {Lynch}, {McEwen}, {Al Noori}, {Ransom}, {Siemens}, \&
  {Surnis}}]{swiggum2023psr}
{Swiggum}, J.~K., {Pleunis}, Z., {Parent}, E., {et~al.} 2023, \apj, 944, 154

\bibitem[{{Swiggum} {et~al.}(2015){Swiggum}, {Rosen}, {McLaughlin}, {Lorimer},
  {Heatherly}, {Lynch}, {Scoles}, {Hockett}, {Filik}, {Marlowe}, {Barlow},
  {Weaver}, {Hilzendeger}, {Ernst}, {Crowley}, {Stone}, {Miller}, {Nunez},
  {Trevino}, {Doehler}, {Cramer}, {Yencsik}, {Thorley}, {Andrews}, {Laws},
  {Wenger}, {Teter}, {Snyder}, {Dittmann}, {Gray}, {Carter}, {McGough},
  {Dydiw}, {Pruett}, {Fink}, \& {Vanderhout}}]{swiggum2015psr}
{Swiggum}, J.~K., {Rosen}, R., {McLaughlin}, M.~A., {et~al.} 2015, \apj, 805,
  156

\bibitem[{{Tauris} {et~al.}(2017){Tauris}, {Kramer}, {Freire}, {Wex}, {Janka},
  {Langer}, {Podsiadlowski}, {Bozzo}, {Chaty}, {Kruckow}, {van den Heuvel},
  {Antoniadis}, {Breton}, \& {Champion}}]{tauris2017formation}
{Tauris}, T.~M., {Kramer}, M., {Freire}, P.~C.~C., {et~al.} 2017, \apj, 846,
  170

\bibitem[{{Tauris} {et~al.}(2015){Tauris}, {Langer}, \&
  {Podsiadlowski}}]{tauris2015ultra}
{Tauris}, T.~M., {Langer}, N., \& {Podsiadlowski}, P. 2015, \mnras, 451, 2123

\bibitem[{{Taylor}(1992)}]{taylor1992pulsar}
{Taylor}, J.~H. 1992, Philos. Trans. Royal Soc. A, 341, 117

\bibitem[{{Taylor} {et~al.}(1979){Taylor}, {Fowler}, \&
  {McCulloch}}]{taylor1979decay}
{Taylor}, J.~H., {Fowler}, L.~A., \& {McCulloch}, P.~M. 1979, \nat, 277, 437

\bibitem[{{Taylor} \& {Weisberg}(1982)}]{taylor1982new}
{Taylor}, J.~H. \& {Weisberg}, J.~M. 1982, \apj, 253, 908

\bibitem[{{Taylor} \& {Weisberg}(1989)}]{taylor1989further}
{Taylor}, J.~H. \& {Weisberg}, J.~M. 1989, \apj, 345, 434

\bibitem[{{van den Heuvel}(2019)}]{van2018high}
{van den Heuvel}, E. P.~J. 2019, IAU Symposium, 346, 1

\bibitem[{{van Leeuwen} {et~al.}(2015){van Leeuwen}, {Kasian}, {Stairs},
  {Lorimer}, {Camilo}, {Chatterjee}, {Cognard}, {Desvignes}, {Freire},
  {Janssen}, {Kramer}, {Lyne}, {Nice}, {Ransom}, {Stappers}, \&
  {Weisberg}}]{van2015binary}
{van Leeuwen}, J., {Kasian}, L., {Stairs}, I.~H., {et~al.} 2015, \apj, 798, 118

\bibitem[{{Vigna-G{\'o}mez} {et~al.}(2018){Vigna-G{\'o}mez}, {Neijssel},
  {Stevenson}, {Barrett}, {Belczynski}, {Justham}, {de Mink}, {M{\"u}ller},
  {Podsiadlowski}, {Renzo}, {Sz{\'e}csi}, \& {Mandel}}]{vigna2018formation}
{Vigna-G{\'o}mez}, A., {Neijssel}, C.~J., {Stevenson}, S., {et~al.} 2018,
  \mnras, 481, 4009

\bibitem[{{Weisberg} \& {Huang}(2016)}]{weisberg2016relativistic}
{Weisberg}, J.~M. \& {Huang}, Y. 2016, \apj, 829, 55

\bibitem[{{Wolszczan}(1991)}]{wolszczan1991nearby}
{Wolszczan}, A. 1991, \nat, 350, 688

\bibitem[{{Yao} {et~al.}(2017){Yao}, {Manchester}, \& {Wang}}]{yao2017new}
{Yao}, J.~M., {Manchester}, R.~N., \& {Wang}, N. 2017, \apj, 835, 29

\end{thebibliography}

\begin{appendix}

\section{Implementation of the DM jump fit}\label{DMjump}

One of the most relevant differences between APSUSE and PTUSE is the implementation of coherent de-dispersion during recording for PTUSE. This leads to a discrepancy between the best DM in the two data sets, as APSUSE data suffers from intra-channel smearing. While this hampers the $S/N$ of APSUSE-derived ToAs, it also introduces a best-DM discrepancy between the APSUSE-derived ToAs and the PTUSE-derived ToAs, which leads to spurious DM trends in the time series when combined, and therefore to biased fit parameters and reduced quality fit. We implement a DM jump between the two data sets to enforce the continuity of the DM variations in the data, and to ensure a good quality fit of the DM and DM1 parameters in the DDH and DDGR models, taking advantage that in some epochs the data sets overlap.

Such DM jump is not implemented in \texttt{TEMPO2}, and therefore we implement a manual fitting instead. The process is as follows: firstly, a global fit is implemented on the entire data set, including DM and DM1. Such fit has an unreliable DM1 value because APSUSE data dominates the beginning of the time series, while PTUSE dominates the middle and later half. Then, we take the resulting model and fit only DM and DM1 for the APSUSE and PTUSE data sets individually. This creates two individual DM models, one for each data set, represented as
\begin{equation}
\begin{aligned}
& \textrm{DM}^\textrm{APSUSE}\left(t\right)=DM_0^\textrm{APSUSE}+t\times DM_\textrm{p}^\textrm{APSUSE}\textrm{ and} \\
& \textrm{DM}^\textrm{PTUSE}\:\:\left(t\right)=DM_0^\textrm{PTUSE}\:\:+t\times DM_\textrm{p}^\textrm{PTUSE}\textrm{,}
\end{aligned}
\end{equation}
where $DM_0^i$ and $DM_\textrm{p}^i$ are the DM value at a reference time and its constant derivative, and $t$ the instantaneous time, which covers a diferent range for each data set. Then, the average of $\Delta\textrm{DM}=\textrm{DM}^\textrm{PTUSE}-\textrm{DM}^\textrm{APSUSE}$ is computed in the overlapping range, and the DM value of the APSUSE template is shifted by $-\Delta\textrm{DM}$. Subsequently, the APSUSE ToAs are re-processed with this fix. This cycle is repeated several times until the $\Delta\textrm{DM}$ value is seen to approach zero. In our implementation, the profiles were de-dispersed originally with $\textrm{DM}=344.312$~pc\,cm$^{-3}$, and the value that sets $\Delta\textrm{DM}\sim0$ was found to be $\textrm{DM}^\textrm{APSUSE}=344.336$~pc\,cm$^{-3}$. %344.3359375

The setup for the DDH and DDGR fits and their measurements are listed in Table~\ref{measurements}. The timing residuals are plotted in Fig.~\ref{residuals}.

\section{Implementation of the MerKAT coherent beam pattern in \texttt{PsrPopPy2}}\label{beamPatternPsrPopPy2}

\begin{figure}
\centering
 \includegraphics[width=\columnwidth]{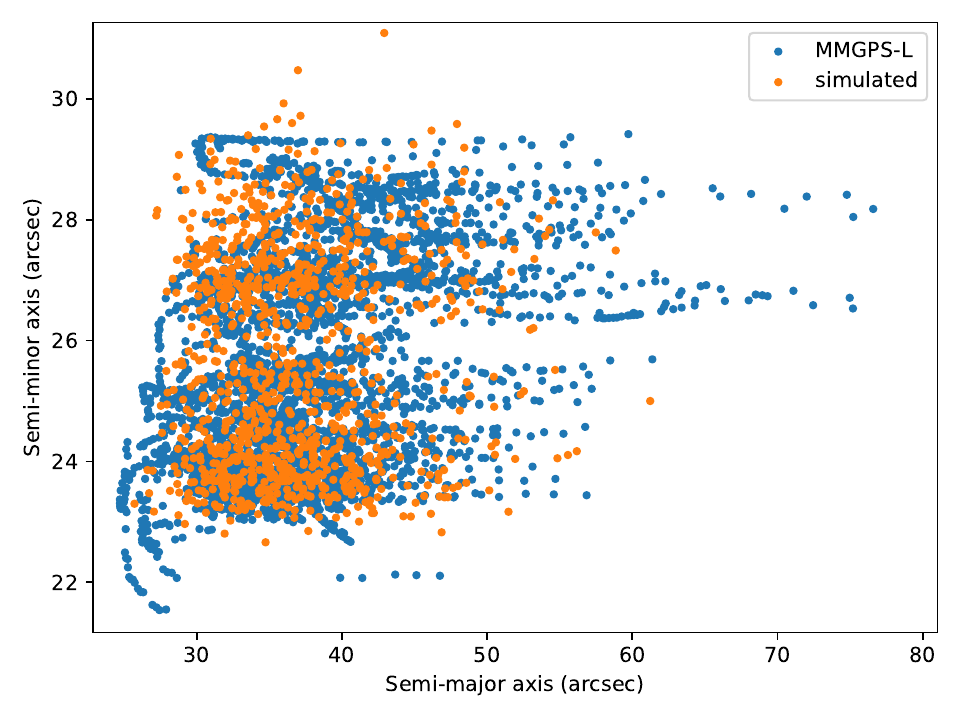}
 \caption{\label{coherent_sizes} Randomly computed coherent beam sizes for the \texttt{PsrPopPy2} simulation of the MMGPS-L survey for the NS-NS merger rate computation (orange) against the true distribution of coherent beam sizes of the bulk of the MMGPS-L pointings (blue).}
\end{figure}

Unlike the pulsar surveys included in the previous NS merger rate estimates, the MMGPS-L makes use of the MeerKAT multi-dish telescope. It is, therefore, necessary to implement the interferometric nature of its observations, which manifest as a tiling of coherently phases beams withing the main survey beam defined by the resolution of individual antennas \citep{chen2021wide,padmanabh2023mmgps-l}.

Each MMGPS-L pointing consists of a survey beam with a Full Width at Half Maximum (FWHM) of 29.85~arcmin, within which a tiling of 480 coherently phased beams is embedded \citep{chen2021wide,padmanabh2023mmgps-l}. The size and orientations of these coherent beams depends on the local sky position and the antenna array configuration, and therefore they can be considered independent of the RA and DEC coordinates \citep{chen2021wide}. Thus, we consider them random in nature, with a probability distribution in semi-major and semi-minor axis that mimics the true distribution of MMGPS-L size values. The probability distributions are derived by fitting skewed Gaussian functions to the full set of MMGPS-L pointing size values, and upon simulation of random values, we see that this method is successful in mimicking typical MMGPS-L coherent beam sizes (Fig.~\ref{coherent_sizes}).

A second aspect of randomness in the simulation of MMGPS-L pointings is the size of the coherent beam tiling itself. Before October 5th, 2021, the coherent tiling was circular and its radius $r_\textrm{c}$ depended on the local sky position \citep{padmanabh2023mmgps-l}. In most of the pointings, it was held that $r_\textrm{c}<\textrm{FWHM}/2$, but occasionally the opposite was true. We find that a sum of two normal Gaussian functions with standard deviations of $\sigma_1=0.898$ and $\sigma_2=2.264$~arcmin, central values of $\langle x_1\rangle=12.546$ and $\langle x_2\rangle=12.761$~arcmin and a height ratio of $A_1/A_2=2.942$ describes the distribution of the distance of the furthest coherent beam from the centre of the tiling, and take it as a distribution for $r_\textrm{c}$. After October 5th, 2021, the coherent beam tiling was instead forced to adopt an hexagonal shape with an inner radius equal to the survey beam to provide a complete coverage of the sky, which allows us to model the tiling consistently across pointings \citep{padmanabh2023mmgps-l}.

This parametrisation is implemented to \texttt{PsrPopPy2} by modifying the \texttt{doSurvey.py} script with the computation of an extra degradation factor $0\leq  D_{coh}\leq1$ for the coherent beams on top of the Gaussian degradation factor of the survey beam $0\leq D_{sur}\leq1$ used for single-dish surveys\footnote{\url{https://github.com/mcbernadich/DNS-merger-rate-2022}}. The process goes as following: for each computed $D_{sur}$, we randomly select whether the pointing should be considered under the older or newer rules for tiling filling (before or after October 5th, 2021) with a 0.39 or 0.61 chance, corresponding to the fraction of pointings under the old and new tiling rules, respectively. In the case that the newer tiling rules are assigned, then we assume an efficient hexagonal tiling which takes an area of
\begin{equation}A=\left(29.85/2~\textrm{arcmin}\right)^2\times6\times\frac{\sin{30^\circ}}{\cos{30^\circ}}\textrm{,}\end{equation}
where the 480 coherent beams are uniformly distributed. Otherwise, the area taken by the tiling is
\begin{equation}A=\pi\times\left(0.95\times r_\textrm{c}~\textrm{arcmin}\right)^2\textrm{,}\end{equation}
where the 480 coherent beams are also uniformly distributed, and where the 0.95 factor comes from the fact that the outermost coherent beam is typically an outlier slightly outside of the main radius. This area is then divided between the 480 beams and a random position for a potential pulsar discovery is drawn within the square of the corresponding size. Then, $D_{coh}$ is a second degradation factor computed from a 2D Gaussian function with the random dimensions drawn from the coherent beam tiling size distributions (Fig.~\ref{coherent_sizes}) centered at the square and aligned with its sides. However, for cases from the older tiling rules in which $r_\textrm{c}<\textrm{FWHM}/2$, if the offset from the centre of the survey beam is found to be larger than $1.05\times r_\textrm{c}$, then we set $D_{coh}=0$ instead.

Finally, like in the previous works \citep{grunthal2021revisiting,pol2020updated}, each pulsar in a DNS merging within the Hubble time has a computed Doppler degeneration factor that depends on the orbital parameters, the spin period, the duration of the observations and the applied search method based on the definitions given by \cite{bagchi2013detectability}. Like the other surveys, the MMGPS-L implements an acceleration search, and therefore we computed the parameter $0\leq \gamma_{2m}\leq1$ from equation (11) in \cite{bagchi2013detectability} using the code\footnote{\url{https://github.com/NihanPol/SNR_degradation_factor_for_BNS_systems}} prepared in \cite{pol2019future}, under the assumption of 500-second long observations. Therefore, the $S/N$ of a simulated pulsar in \texttt{PsrPopPy2} is computed as
\begin{equation}S/N_\textrm{pulsar}=S/N_\textrm{input}\times D_\textrm{sur}\times D_\textrm{coh}\times\gamma_{2m}^2\textrm{.}\end{equation}
For the MMGPS-L, we only consider as discoveries signals with $S/N_\textrm{pulsar}\geq9$.

\end{appendix}

\end{document}